\documentclass[usenatbib]{mnras}

\usepackage{newtxtext,newtxmath}
\usepackage[T1]{fontenc}
\usepackage{ae,aecompl}

\usepackage{graphicx}	% Including figure files
\usepackage{amsmath}	% Advanced maths commands
\usepackage{enumitem}
\usepackage[caption=false]{subfig}
\usepackage{calc}
\usepackage{comment}
\usepackage{threeparttable}
\usepackage{xcolor}
\usepackage{lipsum}
\newcommand{\figurenv}[3][]{\begin{figure}
\centering
\includegraphics[width=\linewidth]{plots/#2}
\caption{#3}
\end{figure}}

\newcommand{\pagefigurenv}[2]{\begin{figure*}
\centering
\includegraphics[width=\linewidth]{plots/#1}
\caption{#2}
\end{figure*}}

\def\rvir/{\ensuremath{R_{\mathrm{vir}}}}
\def\rstarhalf/{\ensuremath{R_{\mathrm{s},1/2}}}
\def\rhalf/{\ensuremath{R_{1/2}}}
\def\ellpatch/{\ensuremath{\ell_\mathrm{patch}}}

\def\VTS/{\ensuremath{\mathrm{VTS}}}

\def\etherm/{\ensuremath{E_{\mathrm{therm}}}}
\def\eR/{\ensuremath{E_R}}
\def\ez/{\ensuremath{E_z}}
\def\ephi/{\ensuremath{E_\phi}}
\def\etot/{\ensuremath{E_{\mathrm{kin+therm}}}}
\def\ekin/{\ensuremath{E_{\mathrm{kin}}}}

\def\tbursty/{\ensuremath{t_{\mathrm{bursty}}}}
\def\zbursty/{\ensuremath{z_{\mathrm{bursty}}}}
\def\tcooltff/{\ensuremath{t_{\mathrm{cool}}^{(s)}/t_\mathrm{ff}}}

\def\tcools/{t_{\rm cool}^{(s)}}
\def\tff{t_{\rm ff}}

\def\mtwelvei/{\texttt{m12i}} %\_res7100
\def\mtwelvef/{\texttt{m12f}} %\_res7100
\def\mtwelveb/{\texttt{m12b}} %\_res7100

\defcitealias{Gurvich2020}{G20}
\def\paperone/{\citetalias{Gurvich2020}}

\newcommand{\northwestern}{Department of Physics \& Astronomy and CIERA, Northwestern University, 1800 Sherman Ave, Evanston, IL 60201, USA}
\newcommand{\telaviv}{School of Physics \& Astronomy, Tel Aviv University, Tel Aviv 69978, Israel}
\newcommand{\caltech}{
TAPIR, Mailcode 350-17, California Institute of Technology, Pasadena, CA 91125, USA}
\newcommand{\ucdavis}{Department of Physics \& Astronomy, University of California, Davis, CA 95616, USA}
\newcommand{\cca}{Center for Computational Astrophysics, Flatiron Institute, 162 5th Ave, New York, NY 10010, USA}
\newcommand{\pomona}{Department of Physics and Astronomy, Pomona College, Claremont, CA, USA}
\newcommand{\durham}{Institute for Computational Cosmology, Department of Physics, Durham University, South Road, Durham DH1 3LE, UK}
\newcommand{\ucirvine}{Department of Physics \& Astronomy, 4129 Reines Hall, University of California, Irvine, CA 92697, USA}

\title[Rapid disc settling in FIRE]{Rapid disc settling and the transition from bursty to steady star formation in Milky Way-mass galaxies}

\author[Gurvich et al.]{
\parbox{\textwidth}{
Alexander B. Gurvich$^{1}$\thanks{E-mail: \href{mailto:agurvich@u.northwestern.edu}{agurvich@u.northwestern.edu}},
Jonathan Stern$^{2}$,  
Claude-Andr\'e Faucher-Gigu\`ere$^{1}$, 
Philip F. Hopkins$^{3}$, 
Andrew Wetzel$^{4}$,
Jorge Moreno$^{5,8,3}$,
Christopher C. Hayward$^{6}$,
Alexander J. Richings$^{7}$,
Zachary Hafen$^{8}$
\vspace{6pt} % lol why do I need to add this??
}
\\
$^{1}${\northwestern}\\
$^{2}${\telaviv}\\
$^{3}${\caltech}\\
$^{4}${\ucdavis}\\
$^{5}${\pomona}\\
$^{6}${\cca}\\
$^{7}${\durham}\\
$^{8}${\ucirvine}
}

\date{Submitted to MNRAS, March 2022}

\pubyear{2021}
\begin{document}
\label{firstpage}
\pagerange{\pageref{firstpage}--\pageref{lastpage}}
\maketitle

\begin{abstract}
Recent observations and simulations indicate substantial evolution in the properties of galaxies with time, wherein rotationally-supported and steady thin discs (like those frequently observed in the local universe) emerge from galaxies that are clumpy, irregular, and have bursty star formation rates (SFRs). 
To better understand the progenitors of local disc galaxies we carry out an analysis of three FIRE-2 simulated galaxies with a mass similar to the Milky Way at redshift $z=0$. 
We show that all three galaxies transition from bursty to steady SFRs at a redshift between $z=0.5$ and $z=0.8$, and that this transition coincides with a rapid (${\lesssim}1$ Gyr) emergence of a rotationally-supported interstellar medium (ISM).
In the late phase with steady SFR, the rotational energy comprises ${\gtrsim}90\%$ of the total kinetic + thermal energy in the ISM, and is roughly half the gravitational energy. 
By contrast, during the early phase with bursty star formation, the ISM has a quasi-spheroidal morphology and its energy budget is dominated by quasi-isotropic flows including turbulence and coherent inflows/outflows. 
This result, that rotational support is subdominant at early times, challenges the common application of equilibrium disc models to the high-redshift progenitors of Milky Way-like galaxies. 
We further find that the formation of a rotation-supported ISM coincides with the formation of a thermal energy-supported inner circumgalactic medium (CGM). 
Before this transition, the inner CGM is also supported by turbulence and coherent flows, indicating that at early times there is no clear boundary between the ISM and inner CGM. 
\end{abstract}

\begin{keywords}
galaxies: formation -- galaxies: evolution -- galaxies: disc -- galaxies: star formation -- galaxies: ISM
\end{keywords}

%%%%%%%%%%%%%%%%%%%%%%%%%%%%%%%%%%%%%%%%%%%%%%%%%%

%%%%%%%%%%%%%%%%% BODY OF PAPER %%%%%%%%%%%%%%%%%%
\section{Introduction}

    Deep surveys of star forming galaxies (SFGs) in the last $10-15$ years have revealed substantial evolution in their morphologies and kinematics with mass and time (see \citealt{FoersterSchreiber2020} for a recent review). 
    SFGs at redshifts $z \gtrsim 1$  are typically observed to have clumpy and/or disturbed morphologies in rest-UV light, which traces recent star formation \citep[e.g.,][]{Elmegreen2007, Law2012, Guo2015}, in contrast with the thin and regular star forming discs commonly observed in the local Universe. 
    Observations of ionized gas kinematics in high-redshift SFGs also often find that the rotational velocity $V_{\rm rot}$ is comparable or subdominant to a disordered velocity component with dispersion $\sigma_{\rm g}$, suggesting that typical local discs ``settle'' from disordered progenitors at $z\lesssim1$ \citep[][]{Kassin2012a, Simons2017}.
    Despite the general decrease in importance of rotation with increasing lookback time, massive dynamically cold discs have been detected as early as $z\approx4$ \citep[][]{Hodge2019,Neeleman2020, Rizzo2020, Rizzo2021}, and irregular star forming morphologies are common also at $z\sim0$ at the low-mass end of the SFG population  \citep[e.g.][]{Roberts69, Hunter97,Simons15}. 
    These observations may indicate that mass, rather than redshift, is a primary determinant of whether a SFG is disky, while the trend with redshift is an indirect result of the growth of mass with time \citep[][]{Wisnioski2015, Harrison2017, Rodrigues2017,Tiley2019,Tiley2021}. 

    Another property of SFGs which appears to evolve with mass and cosmic time is the variability of the star formation rate (SFR).
    Star formation variability can be probed observationally by comparing SFRs inferred using indicators sensitive to different timescales, such as H$\alpha$ which is sensitive to timescales ${\sim}5$ Myr versus far ultraviolet which is sensitive to ${\sim}10-100$ Myr \citep[e.g.][]{Kennicutt2012, Dominguez2015, FloresVelazquez2021}.
    Using this approach several studies have reported evidence that dwarf galaxies have highly time-variable (``bursty'') SFRs which can fluctuate by an order-of-magnitude or more on timescales of $30-100$ Myr, while more massive disc galaxies have more time-steady SFRs \citep[][]{Weisz2012, Kauffmann2014,Sparre2017, Emami2019}. 
    It seems plausible that SF burstiness is connected to a disordered morphology (i.e., a lack of a stable rotating disc) at low mass or high redshift, either if a stable disc is a precondition for regulating star formation as in standard disc equilibrium models \citep[e.g.][]{Faucher-Giguere2013,Krumholz2018}, or if enhanced stellar feedback from strong supernova clustering when star formation is bursty is able to disrupt the nascent disc \cite[e.g.,][]{Martizzi20}.

    Analysing cosmological simulations that reproduce the observational trends mentioned above provides a promising route to understand the physical origin of these trends and the nature of bursty, dispersion-dominated SFGs at high-redshift and low mass.  
    In the simulations one can infer properties such as the timescale of the disc settling process, which are inaccessible to SFG population studies where only a single snapshot per galaxy is available. 

    In this paper we analyse the physics of disc settling and its relation to the transition from bursty to steady star formation using  cosmological zoom-in simulations from the FIRE project \citep[][]{Hopkins2014, Hopkins2018}\footnote{See the FIRE project website: \href{http://fire.northwestern.edu}{http://fire.northwestern.edu}.}. 
    Similar to the trends suggested by observations, early low-mass FIRE galaxies have highly dynamic and clumpy gas morphologies \cite[][]{Hopkins2014, Ma2017}, and during this phase the SFR is order-of-magnitude time variable \citep[][]{Sparre2017,Faucher-Giguere2018,FloresVelazquez2021} while the corresponding bursty stellar feedback drives ``gusty'' galactic winds \citep{Muratov2015, Muratov2017, AA17_cycle, Pandya2021}. 
    At late times (or higher masses) the simulated galaxies develop a long-lived, thin disc similar to local spiral galaxies.  
    
    Previous physical explanations for the observed trends in SFGs typically fall within two broad categories. 
    In the first category some or all of these trends are attributed to a change in the properties of galaxy discs, such as an increase with redshift in the gas fraction of the disc, which makes the disc morphologically thicker and clumpier due to a larger Toomre mass \citep[e.g.,][]{Dekel2009b,Krumholz2018}. 
    A challenge for these type of explanations is that they assume an equilibrium disc exists to begin with, a condition which is not clearly met at high redshift and low mass where many SFGs have $V_{\rm rot}\lesssim\sigma_g$. 
    A second category of explanations attributes observed SFG trends to a change in the rate or effect of mergers with mass and time  \citep[e.g.,][]{Brook2004,Bird2013,Dekel2020a}. 
    It however remains to be seen if sufficiently massive mergers are common enough to cause such a large fraction of low-mass / high-redshift SFGs to appear disturbed. 
    Moreover, in the FIRE simulations we analyse in this paper, the transition from bursty to steady SFR appears too sharp (see \S \ref{s:star_formation}) to be solely attributable to an evolution in the merger rate, since the latter is expected to change substantially only on cosmological time-scales.
    
    There is also a third possibility, recently suggested by \cite{Stern2020} and followed up by \cite{Yu2021} and \cite{Hafen2022}, in which observed SFG trends are driven by a change in the physics of gas surrounding the galaxy, known as the circumgalactic medium (CGM). 
    Due to the increase in gas cooling time in the CGM with mass, the CGM is expected to `virialize' at a mass comparable to that in which discs are observed to settle, transitioning from a relatively cool and dynamic CGM at low masses to a hot and quasi-static CGM at high masses \citep{WhiteRees78,birnboim2003,Keres05,Faucher-Giguere2011a, VandeVoort2012a, vandevoort2016, Fielding2017,Stern2020_hot_max}. 
    This expected transition in the CGM has been shown to occur in the FIRE simulations at small halo radii where cooling times are short relative to dynamical and cosmological timescales \citep{Stern2020}. 
    CGM virialization affects both the physics of inflows onto the galaxy and the physics of galaxy outflows, and hence is likely to have significant effects on the properties of SFGs. In this work we further demonstrate the connection between the thermodynamics of the inner CGM and the properties of SFGs, by showing that the energetics of the ISM prior to disc settling in FIRE are similar to those in the inner CGM prior to virialization, while after the transition these two media are clearly separable. 

    Our analysis builds on \citeauthor{Gurvich2020} (\citeyear{Gurvich2020}, hereafter \paperone/), which focused on low-redshift, Milky Way-mass FIRE galaxies, well after discs have settled and when the SFRs have become steady.  
    In that paper, we developed an analysis framework to quantify the different forms of ISM pressure in different thermodynamic phases.
    Here we analyse the same three simulated galaxies as in \paperone/ but now include the full evolution from early times to $z=0$ using a modified version of the analysis framework that is applicable in both phases.

    The outline of this paper is as follows. 
    In \S \ref{s:sim_methods} we summarize the simulations we analyse, define the coordinate frame and how we attribute gas resolution elements to the ISM or the CGM. 
    In \S \ref{s:properties} we quantify the evolution of basic properties of the main galaxies, including the SFR, gas fraction, disc radius and thickness, and measures of rotational support. 
    In \S \ref{s:energy_metrics} we extend the analysis of \paperone/ to high redshift using an analogous energy-based framework and quantify how energy in the ISM and inner CGM is partitioned between different forms of energy and thermodynamic phases, and how these results change over cosmic time.
    We discuss our results in \S \ref{s:discussion} and summarize our main conclusions in \S \ref{s:conclusion}.

\pagefigurenv{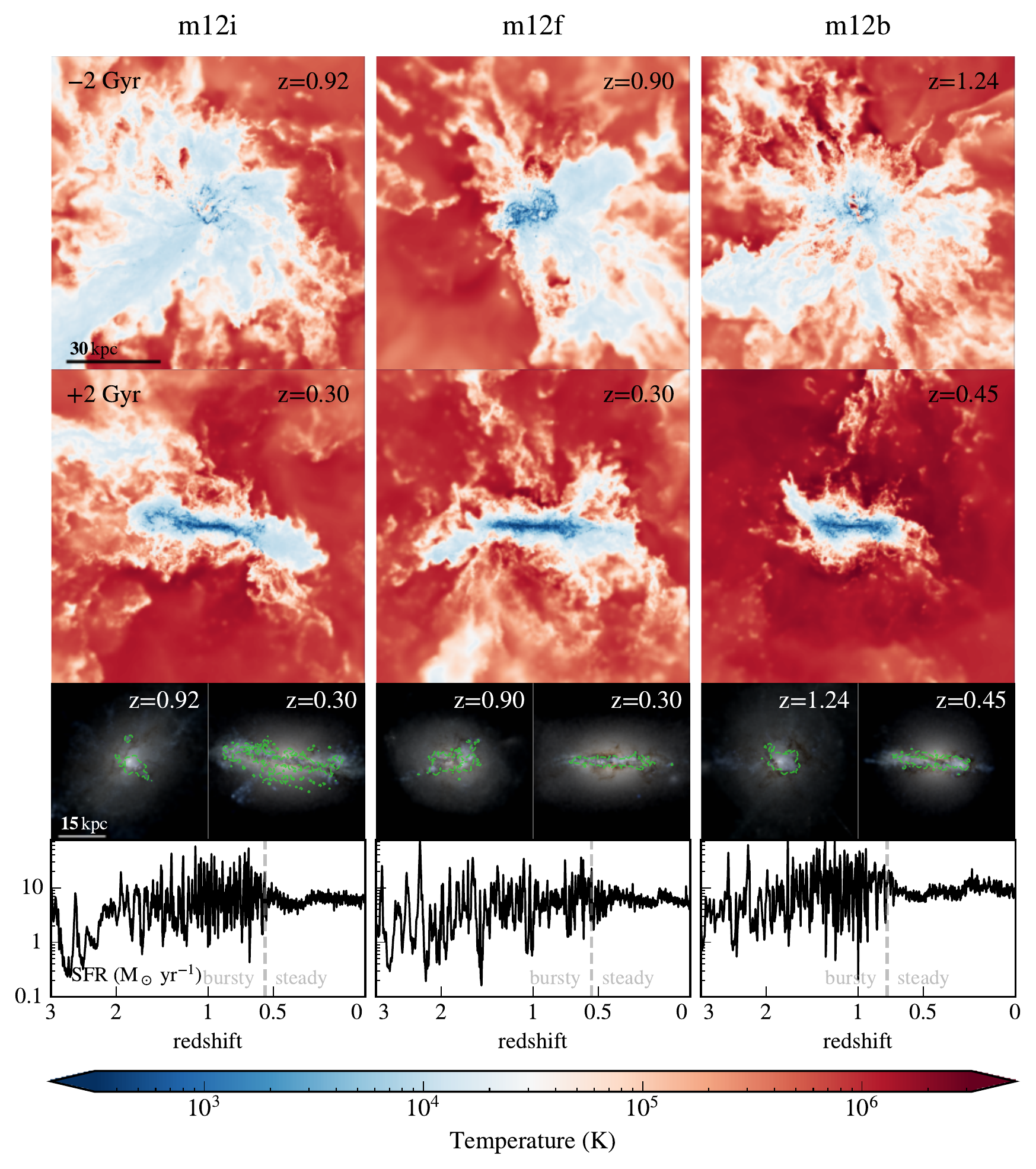}{
    \label{f:galaxy_stamps} Edge-on temperature projections of three Milky Way-mass simulations at 2 Gyr before (top row) and at 2 Gyr after (second row) they undergo a transition from bursty to time-steady SFRs (bottom row).
    The third row shows mock Hubble renderings of simulated starlight for each of the two epochs.
    Green contours contain 99\% of stellar mass younger than 25 Myr, and demonstrate the geometry of the star-forming region of the galaxy.
    Between these two epochs the galaxy transforms from a clumpy, disturbed morphology with a bursty star formation rate, to a thin disc with a time-steady star formation rate.
    The temperature projections also show that during the bursty phase cold, warm, and hot gas mix at large radii ($\gtrsim 30$ kpc) whereas after the transition cold gas forms a disc embedded in  a volume-filling hot galaxy atmosphere. 
    The gas projections and mock Hubble renderings in this figure were generated using FIRE Studio, an open-source visualization software for cosmological simulations \citep{Gurvich:2022}.}
    
\section{Simulations and analysis methods} \label{s:sim_methods}
\subsection{Summary of simulations and their physics}
    
    We use FIRE-2 cosmological zoom-in simulations run with GIZMO, a publicly available gravity+magnetohydrodynamic code\footnote{Information about GIZMO and a public version of the code is available at: \url{https://www.tapir.caltech.edu/~phopkins/Site/GIZMO}}, in its Meshless Finite Mass mode \citep{Hopkins2018}. 
    MFM is a Lagrangian method that combines the advantages of traditional smooth particle hydrodynamics (SPH) and grid-based methods
    \citep[for numerical details and tests, see][]{Hopkins2015}. 
    We provide a short summary of the relevant aspects here, while the full FIRE-2 physics model is described in \citet{Hopkins2018}. 
    
    The simulations include radiative cooling for gas down to $10$ K, including an approximate treatment of fine-structure metal and molecular lines. 
    The gas is irradiated by a uniform, redshift-dependent ionizing background \citep[][]{FG09}, as well as local sources.  
    Star formation occurs in gas that is sufficiently dense ($n_\mathrm{H}\geq 1000$ cm$^{-3}$), self-gravitating, and self-shielding.
    When all criteria for star formation are satisfied, the gas particle is converted into a star particle representing a simple stellar population (that inherits its parent gas particle's mass and metallicity) with 100\% efficiency per local free-fall time. 
    A much lower star formation efficiency on galactic scales, and thus agreement with the observed Kennicutt-Schmidt relation, emerges as a result of regulation by stellar feedback \citep[][]{Hopkins2014, Orr2018}.
    
    The simulations include a multi-channel model for stellar feedback. 
    Star particles return mass, metals, momentum, and energy into the surrounding ISM, using rates that are functions of the star particle's age following the \texttt{STARBURST99} population synthesis model \citep{Leitherer1999}. 
    These feedback processes include supernovae (Type II and Ia), stellar winds from O, B, and AGB stars, photoelectric heating, photoionization, and radiation pressure. 
    The model for radiation pressure includes both short-range and long-range components \citep[][]{Hopkins2020}. 

    In this paper we focus on three representative Milky Way-mass galaxies from the FIRE project, which are chosen to match the sample analysed in \paperone/: \mtwelvei/, \mtwelvef/, and \mtwelveb/ \citep[see][for more details about these simulations]{Wetzel2016,Garrison-Kimmel2017,Garrison-Kimmel2019}.
    We choose to focus on a relatively small number of runs compared to the entire FIRE catalog in order to develop a new set of diagnostics which we intend to apply to the larger sample of FIRE galaxies,  including at different mass scales, in future work.
    The runs we focus on in this paper are standard, hydrodynamic FIRE-2 simulations which include a subgrid model for turbulent metal diffusion but do not include the effects of magnetic fields or cosmic rays. 
    The initial mass resolution in these simulations is $m_\mathrm{b}\approx 7,100$ M$_\odot$ for gas cells and star particles (evolved star particles can have significantly lower mass owing to stellar mass loss). 
    Thus, GMCs with masses $M_{\rm GMC}\sim 10^{6}-10^{7}$ M$_{\odot}$ contain $\gtrsim 100-1000$ resolution elements.
    Gravitational softening lengths for gas cells are adaptive and reach values of ${\sim}5$ pc at the average density of star-forming gas \citep[see Table 3 of][]{Hopkins2018}.

    The simulated Milky-Way mass galaxies all experience a transition from bursty to steady star formation, and form large, long-lived gas discs starting at redshifts $z<1$ \citep[][]{Ma2017,Garrison-Kimmel2018}. 
    Figure~\ref{f:galaxy_stamps} shows images of the gas and stellar distributions centered on the main galaxies before and after disc settling, along with the corresponding star formation histories. 
    The $z=0$ halo masses (all in the range $M_{\rm h}\sim (1-1.5)\times10^{12}$ M$_{\odot}$) and stellar masses are catalogued in Table \ref{t:sim_table} for the simulations we analyse.
    We also list in the table the same quantities at the time when bursty star formation is identified to end, as discussed in \S \ref{s:star_formation} below. 

    In Appendix \ref{a:cosmic_ray} we present results for re-runs of \mtwelvei/ and \mtwelveb/ which include magnetic fields and cosmic-ray feedback from stars. 
    These results indicate that our main conclusions for the ISM are not sensitive to the inclusion of these additional physics.

\subsection{Galaxy tracking and coordinate system}
    As in \paperone/, we use the Amiga Halo Finder \citep[AHF;][]{Gill2004,Knollmann2009} to identify the peak mass density and virial radius $R_\mathrm{vir}$ of the main halo in each zoom-in simulation over cosmic time. 
    The virial radius and virial halo mass are defined using the redshift-dependent overdensity criterion of \citet{Bryan1998}. 
    At each snapshot, we define an analysis coordinate system such that the $z$-axis is parallel to the angular momentum of the stars within 5 \rstarhalf/, where \rstarhalf/ is the radius enclosing half of all the stellar mass within 20\% of \rvir/.
    The origin of the coordinate system ($z=0$) coincides with the halo center (as identified by AHF) and $\phi$ denotes the azimuthal angular coordinate. 
    We have verified that this produces a steadily evolving reference frame that changes from snapshot-to-snapshot (time spacing $\sim 20$ Myr) typically on the order of a degree or less in orientation. 
    \citet{Santistevan2021} suggested that the orientation of the stellar disc stabilizes after the last major merger.
    
    \begin{table}
    \caption{\label{t:sim_table} Simulated galaxy properties.}
    \centering
    \begin{threeparttable}
        \centering
        %% pad with white space so it fills up the column
        \begin{tabular}{c|c|c|c|c|c|c}
          & \multicolumn{2}{c}{$z=0$ } & & & \multicolumn{2}{c}{$z=\zbursty/$}\\ 
          Name & 
          $M_{\mathrm{h}}$ \tnote{a}& 
          $M_{\mathrm{s}}$ \tnote{b}& 
          \tbursty/\tnote{c} & 
          $\zbursty/$ \tnote{c} &
          $M_{\mathrm{h}}$ \tnote{a} & 
          $M_{\mathrm{s}}$ \tnote{b}\\
          \hline
m12i & 1.18 & 6.36 & 8.34 & 0.56 & 0.81 & 3.41\\
m12f & 1.62 & 6.78 & 8.41 & 0.55 & 1.07 & 3.41\\
m12b & 1.30 & 9.80 & 7.11 & 0.76 & 0.76 & 3.69\\ \hline
        \end{tabular}
        \begin{tablenotes}
            \item[a] Total mass (dark matter+baryon) within $R_\mathrm{vir}$  in units of $10^{12}$ M$_\odot$. 
            \item[b] Total stellar mass within 5\rstarhalf/ in units of $10^{10}$ M$_\odot$.
            \item[c] Cosmic time (in Gyr) and redshift of the transition from bursty to steady star formation.
        \end{tablenotes}
        \end{threeparttable}
\end{table}

\subsection{Definition of ISM and inner CGM}

    We show below that at early times in our simulations gas at galaxy radii is not supported by rotation, and its geometry is quasi-spherical rather than disc-like. 
    Thus, in order to use a consistent definition of the ISM at all times (i.e. before and after the emergence of a stable galactic disc), we estimate ISM properties using mass-weighted averages on gas within a spherical radius $r<0.05\rvir/$.
    Our main conclusions below are unchanged if instead we define the ISM using the more complicated cylindrical cut applied in \paperone/, where the ISM is defined as gas within a radial-dependent disc height $|z|<h(R)$. 
    The similarity of the spherical and cylindrical cuts follows from the fact that at late times, when a disc has formed, the spherical averages are dominated by gas within the disc plane, while at early times $h(R)\sim R$.
    
    The choice of maximum ISM radius $r=0.05\rvir/$ is motivated by the expected radius of angular momentum support: gas with a spin parameter equal to the average value of dark matter haloes $\lambda=0.035$  \citep[e.g.,][]{rodriguezpuebla16} will be angular momentum supported\footnote{This follows from the spin parameter definition $j=\sqrt{2}\lambda\rvir/v_{\rm vir}$ \citep{Bullock01}, so for a potential which is roughly isothermal we get $j=v_c r$ at $r=\sqrt{2}\lambda\rvir/$.} within $\sqrt{2}\lambda\rvir/\approx0.05\rvir/$. 
    The gaseous discs which form at late times in our simulations indeed extend roughly out to this radius, supporting this choice (see below and Figure~13 in \citealt{Stern2020}). 
    The properties of the `inner CGM' are estimated using mass-weighted averages of gas within a shell at $0.1\rvir/$ and with width of $0.01\rvir/$. 
    The radius $0.1\rvir/$ is the smallest radius where angular momentum support does not significantly affect spherical averages at all times in our simulations, and can thus be considered `CGM'.  
    The sensitivity of our conclusions to these radial choices is discussed below in \S \ref{s:ism_cgm_distinction}. 
    
\section{Basic properties of the galaxies over cosmic time} \label{s:properties}

    \subsection{The transition from bursty to steady star formation}\label{s:star_formation}
    
\figurenv{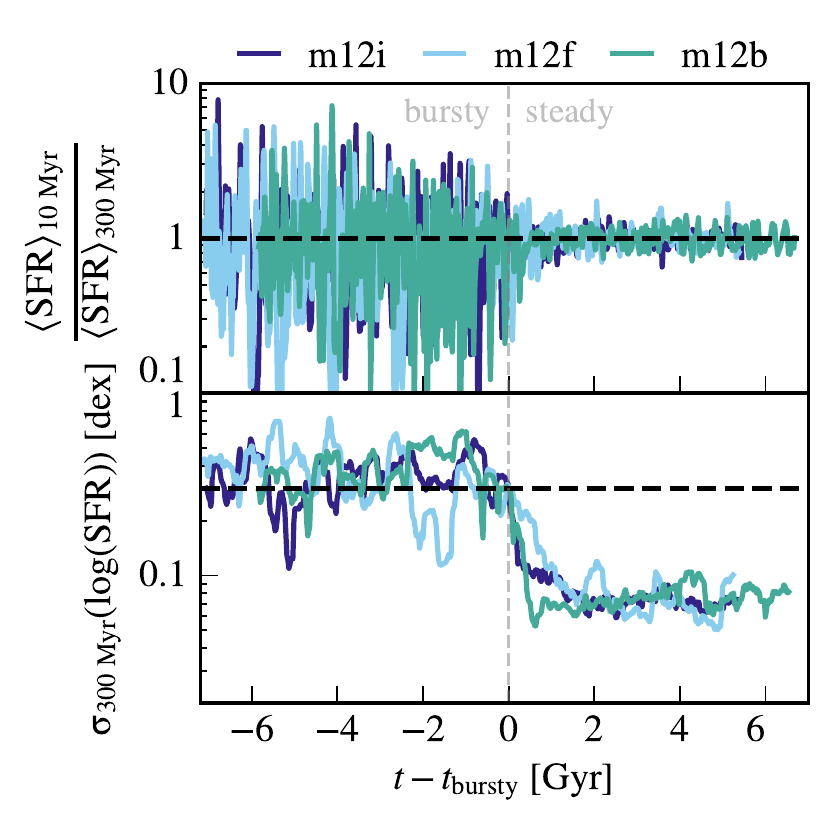}{
    \label{f:star_formation_history} 
    \textit{Top:} The star formation history of three Milky Way-mass simulations normalized by the running average over 300 Myr and offset in time by the identified end of bursty star formation, \tbursty/.
    \textit{Bottom:} The running scatter of $\log\mathrm{SFR}$ in 300 Myr windows.
    The transition epoch \tbursty/ is defined as the time after which this scatter stays below $0.3$ dex.
    When offset by \tbursty/, the three different simulations nearly overlap in both their normalized SFH and running SFH scatter.} 
    
        We measure the star formation histories (SFHs) of each simulation using the ``archaeological '' method, following the procedure outlined in \citet{FloresVelazquez2021}. 
        Put briefly, we take the stars within $5\rstarhalf/$ at redshift $z=0$ and account for stellar mass loss according to their ages and the rates from \texttt{STARBURST99} \citep{Leitherer1999}.
        To construct a SFH, we group the stars according to their stellar formation times in bins spaced by 1 Myr.
        These SFHs are shown in the bottom row of Figure~\ref{f:galaxy_stamps} and in the top panel of Figure~\ref{f:star_formation_history}, the latter normalized by a running mean within a $300$ Myr window. 
        The Figure shows that the fluctuations from the mean SFR are large at early times (i.e. star formation is ``bursty'') while at late times the fluctuations from the mean are much smaller (i.e. star formation is ``steady'').
    
        To quantify the fluctuations, we plot the running scatter in the (log base 10 of the) SFH in 300 Myr windows 
        \begin{align}
            &\sigma_\mathrm{300~Myr}\left(\log(\mathrm{SFR})\right) =\notag \\ 
            &\sqrt{
            \frac{\left\langle \log(\mathrm{SFR})^2 \right\rangle_\mathrm{300~Myr} -\langle \log(\mathrm{SFR}) \rangle_\mathrm{300~Myr}^2 }
            {300}}
        \end{align}
        in the bottom panel of Figure~\ref{f:star_formation_history}.
        The denominator $300$ corresponds to the number of SFR samples (spaced by 1 Myr) in the $300$ Myr window. 
        We define \tbursty/ as the time after which $\sigma_\mathrm{300~Myr}$ remains below $0.3$ dex (${\approx}$ a factor of 2).
        Our main results are not sensitive to the exact threshold used because $\sigma_\mathrm{300~Myr}$ drops rapidly by about a factor of five within a Gyr or less from $t_{\rm bursty}$.
        We also considered two additional definitions based on the peak-to-trough ratio in moving windows and the ratio of the running scatter to the running average in linear SFR as in \cite{Yu2021}, and found variations in \tbursty/ of a few hundreds of Myr.
        This uncertainty in \tbursty/ is short relative to cosmological timescales of $5-10$ Gyr at the transition epoch. 
        In Figure~\ref{f:star_formation_history} and following figures we plot quantities as a function of $t-\tbursty/$ rather than $t$, in order to highlight the correspondence between the transition from bursty to steady star formation, disc settling, and other changes in galaxy properties. 

\figurenv[,height=1.5\linewidth]{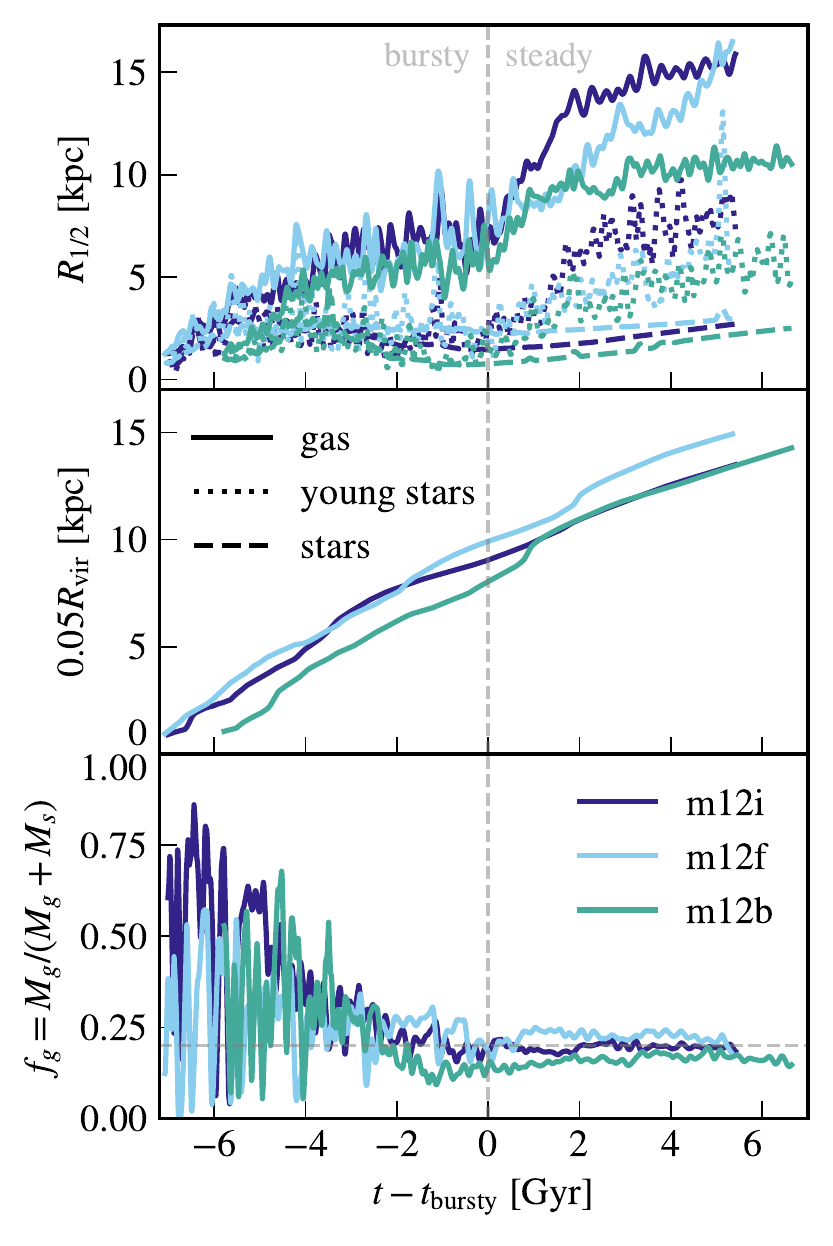}{
    \label{f:basic_properties} 
    Properties of the galaxies in our sample as a function of cosmic time, offset by \tbursty/ (defined in the bottom panel of Figure~\ref{f:star_formation_history}) and smoothed in a moving $100$ Myr window for readability.
    \textit{Top:} the cylindrical radius that encloses half of the mass in each of the gas (solid), stars (dashed), and young stars (${\leq}25$ Myr old; dotted). 
    During the bursty phase, the young and old stars have the same (relatively small) extent. 
    In the steady phase the young star and gas distributions become more extended.
    \textit{Bottom:} the baryonic gas fraction, $f_{\rm g}=M_\mathrm{g}/(M_\mathrm{g}+M_\mathrm{s})$, within $0.05\rvir/$.
    Near the SFR transition the gas fraction settles on a value of $\approx0.2$.}
    
    \subsection{Galaxy radii and gas fractions} \label{s:galaxy_basics}
    
        In the top row of Figure~\ref{f:basic_properties} we show the cylindrical half-mass radius for the gas, stars, and young stars (those with age ${\leq} 25$ Myr) within $0.1\rvir/$, smoothed with a moving $100$ Myr window to improve readability. 
        Before \tbursty/, the young and total stellar populations are concentrated at relatively small radii of a few kpc.  
        After \tbursty/, the gas distribution expands and active star formation extends to larger median radii of $5-10$ kpc.
        The amount of star formation at these larger radii is however small compared to the integrated star formation during the bursty phase and as a result the half-mass radius of the total stellar population remains similar to that in the bursty phase.
    
        In the bottom panel of Figure~\ref{f:basic_properties} we plot the evolution of the baryonic gas mass fraction $f_{\rm g}=M_{\rm g}/(M_{\rm g}+M_{\rm s})$ within a sphere of radius $0.05\rvir/$, again smoothed using a moving $100$ Myr window. 
        The gas fraction is an important quantity in many equilibrium disc models, since the disc thickness and the characteristic mass of star-forming clumps (the Toomre mass) scale with the gas fractions \citep[e.g.,][]{Thompson2005, Dekel2009b, Faucher-Giguere2013, Hayward2017,Faucher-Giguere2018}. 
        The panel shows that the gas fraction decreases steadily from large values ($f_\mathrm{g} \gtrsim 50-60\%$) at high redshift to a more modest value of ${\approx} 20\%$  at ${\sim}\tbursty/$, where the gas fractions plateau at values similar to local spiral galaxies \citep[e.g.,][]{Leroy2008}.

    \subsection{Geometric/kinematic metrics of disc settling} \label{s:disk_settling}
        In order to understand the relation between the transition from bursty to steady star formation and disc settling, we analyse the evolution of multiple metrics of ``diskiness'' based on geometrical properties, kinematics, and angular momentum.
        Three such metrics of diskiness are shown in Figure~\ref{f:diskiness}, as a function of time relative to the identified transition from bursty to steady star formation, $t_{\rm bursty}$.
        As for Figure~\ref{f:basic_properties}, we smooth the curves using a moving 100 Myr window for improved readability.
    
\figurenv[height=1.5\linewidth]{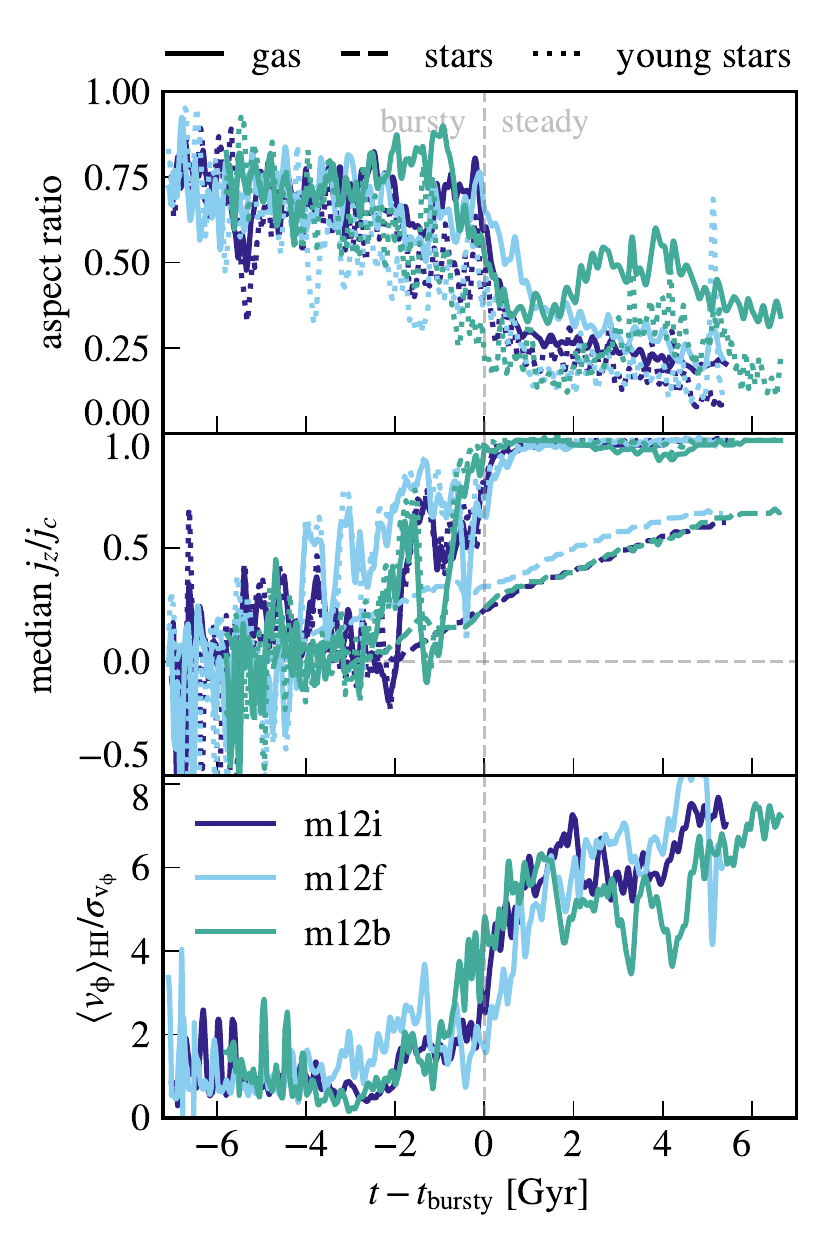}{
    \label{f:diskiness}
    Three metrics of ``diskiness'' in our simulated galaxies versus cosmic time offset by \tbursty/, and smoothed in a moving 100 Myr window for readability.
    \textit{Top:} morphological axis aspect ratio (see \S \ref{s:disk_settling} for definition), for young stars (dotted) and for gas within 0.1 \rvir/  (solid).  
    \textit{Middle:} the median circularity $j_z/j_c$, defined as the ratio of specific angular momentum in the $z$-direction to the specific angular momentum of gas in circular orbits, for gas within 0.1 \rvir/. 
    \textit{Bottom:} the ratio of the HI-weighted average of the rotational velocity to the HI-weighted dispersion in the rotational velocity.
    All panels suggest the appearance of a thin star-forming gaseous disc at $t\sim \tbursty/$.\vspace{-12pt}}
        
        The top panel in Figure~\ref{f:diskiness} plots axis aspect ratios for the gas (solid) and for young stars within 0.1 \rvir/. 
        For a quasi-spheroid this ratio is expected to be ${\sim} 1$, while it is expected to be ${\lesssim}0.2$ for a thin disc similar to those in observed in the local Universe \citep[e.g.,][]{Padilla08}. 
        The axial ratio is measured by dividing the smallest principal axis with the mean of the two remaining principal axes, where the principal axes' lengths are calculated from the eigenvectors and eigenvalues of the moment of inertia tensor. 
        These eigenvectors and eigenvalues define an ellipsoid approximating the orientation and extent of the mass distribution.
        This approach has the advantage of not relying on fitting a parametric model (e.g.\ an exponential profile or a cylindrical disc) which may not be applicable during the bursty phase. 
        Hot gas with $T>10^5$ K is excluded because it can trace the virialized CGM (if applicable) or thermalized outflows, rather than the ISM. 
        The figure shows that at early times, the axis aspect ratio is $0.6-0.8$ for both the young stars and the gas, indicating a geometry closer to spherical than disc-like. 
        At late times, the aspect ratios decrease to typically ${\lesssim}0.3$, indicating a more disc-like geometry. 

    In the middle panel of Figure~\ref{f:diskiness} we plot the median (among resolution elements) specific angular momentum in the $z$-direction, $j_z$, normalized by the specific angular momentum of a circular orbit, $j_c\equiv v_c r$, for the gas and stars within $0.05$ \rvir/ ($v_c$ is the circular velocity). 
    The ratio $j_z/j_c$ is bounded between $-1- 1$, corresponding to a circular orbit that is aligned or anti-aligned with the orientation of the coordinate system. 
    This ratio is sometimes referred to as the ``circularity'', and circularities above 0.8 are interpreted as belonging to a ``thin-disc'' \citep[see for example][]{Yu2021}.
    For both the gas and young stars the median $j_z/j_c$  increases from ${\approx}0$ with a large dispersion to uniformly $\approx1$ in the $\lesssim 2$ Gyr leading up to \tbursty/, indicating the formation of a thin disc. 
    
    In the third panel of Figure~\ref{f:diskiness}, we consider the ratio of the rotational velocity to the velocity dispersion in the neutral gas. 
    Specifically, we compute mean rotational velocity (in the $\phi$-direction) weighted by HI mass,  $\langle v_{\phi} \rangle_{\rm HI}$, including all gas inside $0.05R_{\rm vir}$. 
    We also compute the standard deviation of $v_{\phi}$ for the same gas, $\sigma_{v_{\phi}}$, also weighted by HI mass, and the ratio $\langle v_{\phi} \rangle_{\rm HI}/\sigma_{v_{\phi}}$. 
    This kinematic ratio is useful because it is close to how the degree of rotational support is often estimated in observations \citep{ElBadry2018_HI}.
    This metric as well shows a strong increase from $\approx1$ roughly two Gyr before \tbursty/ to values of $\approx6$ a Gyr after \tbursty/.

    The three metrics of diskiness plotted in Figure~\ref{f:diskiness} all indicate the emergence of a thin gas disc around \tbursty/, with an increase in diskiness appearing to begin $1-2$ Gyr prior to \tbursty/. 
    In the next section we examine these trends in more detail using energy-based metrics. 
    
\section{Energy-based metrics of disc settling} \label{s:energy_metrics}
    We now present an energy-based analysis of how the gas evolves in the simulations which is analogous to the pressure-based analysis in \paperone/ for the same simulated galaxies in the late-time, steady phase. 
    We present these results in terms of energy, rather than pressure (which is proportional to energy density), because the evolution from the early bursty phase to the late steady phase is expressed more clearly this way. 
    For example, we can straightforwardly interpret the decomposition of the total kinetic energy along different cylindrical coordinate axes which, as we show below, is a useful way to quantify disc settling.
    In \paperone/, we also quantified the balance between ISM pressure and the weight of the overlying gas in the disky ISM, which we referred to as ``vertical pressure balance''. 
    In this paper, we instead define an analogous metric, the ratio of the gravitational binding energy of gas to its total (kinetic+thermal) energy, that can be applied equally well before and after disc settling to establish whether the ISM is in some similar state of equilibrium. 

\figurenv{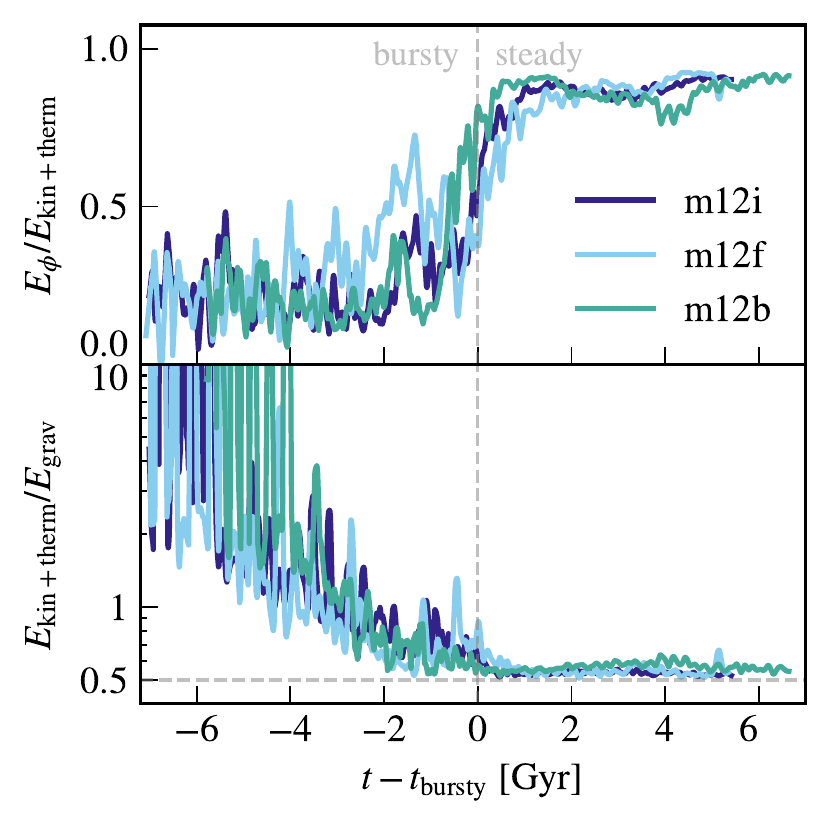}{
    \label{f:energy_summary} 
    \textit{Top:} Ratio of rotational energy to total kinetic+thermal energy in gas at $r<0.05\rvir/$, versus cosmic time offset by \tbursty/ for the three simulations.
    \textit{Bottom:} The ratio of the total kinetic+thermal gas energy to the gravitational potential energy at $<0.05\rvir/$.
    The horizontal dashed line marks the ratio for a cold, rotationally-supported disc. 
    In all simulations a rotation-dominated gaseous disc is apparent only from \tbursty/ onward. }

\pagefigurenv{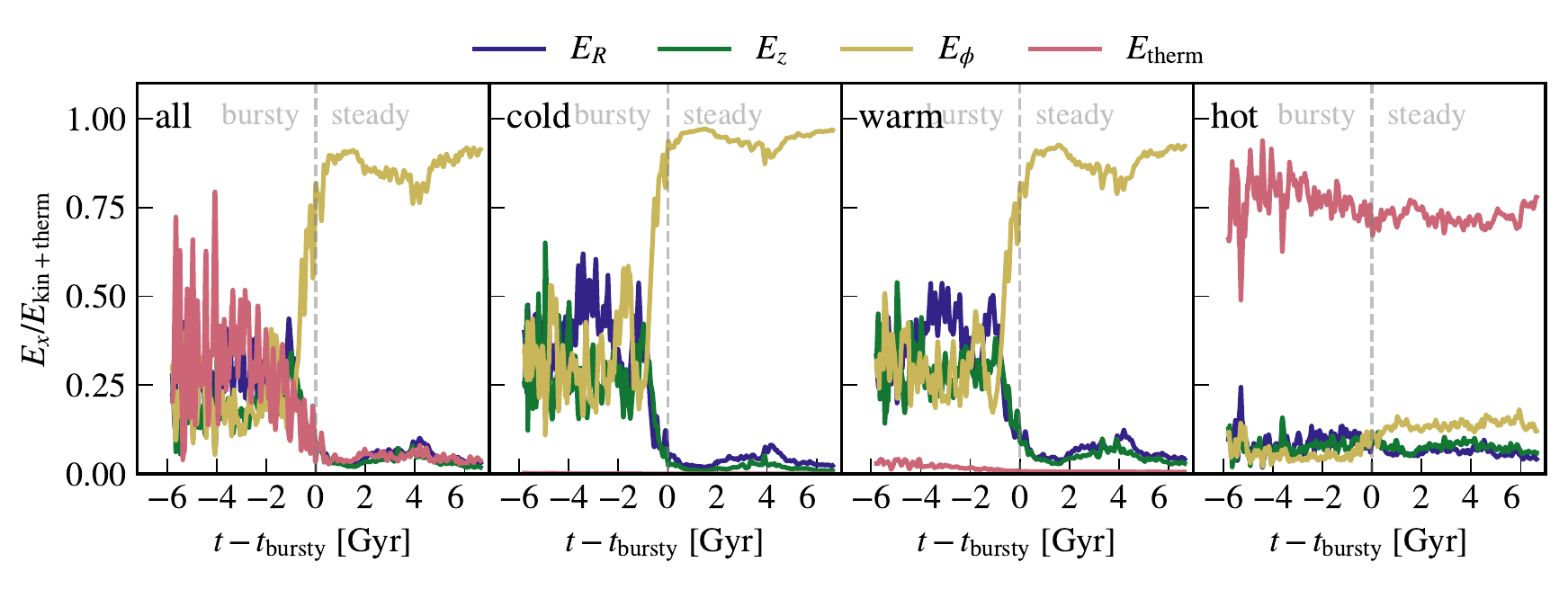}{
    \label{f:type_breakdown}
     A partitioning of the total gas energy within $0.05 \rvir/$ versus cosmic time offset by \tbursty/, for \mtwelveb/. 
     Different colors denote the thermal and three kinetic components (in cylindrical coordinates). 
     The left panel shows all gas, while the three right panels focus on specific thermal phases: cold ($T < 1000$ K, 2$^{\rm nd}$ panel), warm ($1000 < T , 10^{5}$ K, 3$^{\rm rd}$ panel) and hot ($T>10^5$ K, right-most panel). 
     A rotation-dominated ISM emerges only at $t\sim \tbursty/$. 
     At earlier times the ISM energy is distributed roughly equally between the different kinetic and thermal components, i.e.~rotational support is subdominant in the ISM. }

\pagefigurenv{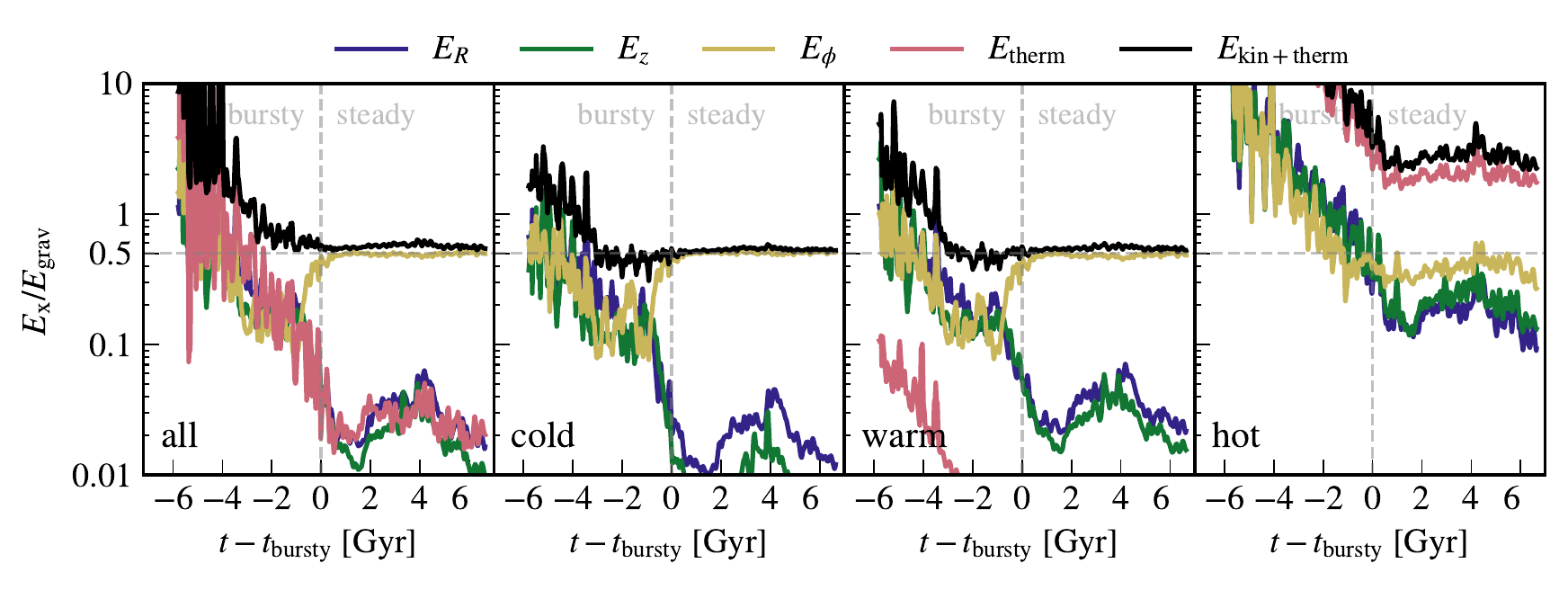}{
    \label{f:grav_ratio_type_breakdown}
    The ratio of different energy components to the gravitational energy, in gas within $0.05 \rvir/$ in \mtwelveb/. 
    The left panel shows all gas while the three right panels focus on gas in different gas temperature phases, as in Figure \ref{f:type_breakdown}. 
    Rotational energy dominates after \tbursty/ and is then roughly equal to $0.5E_\mathrm{grav}$, as expected for a rotationally-supported, cold disc. 
    Prior to \tbursty/, the total energy exceeds $0.5E_\mathrm{grav}$ with a large dispersion mainly at early times, suggesting the galaxy is in a highly-dynamical state (with some of the gas unbound) rather than near equilibrium virial equilibrium.
    }

    \subsection{Comparison of different forms of energy in the ISM} \label{s:energy_forms}
        In this section we study how disc settling is expressed in the energetics of the ISM. 
        To this end, for each snapshot we calculate the mass $M$, thermal  energy $E_{\rm therm}$, kinetic energy in each of the cylindrical coordinate velocity components, $E_R$, $E_\phi$, and $E_z$, and gravitational potential energy $E_{\rm grav}$ for gas within $0.05\rvir/$. 
        Gas denser than $50$ cm$^{-3}$ is excluded from the analysis, as it is primarily found in dense gravitationally-bound clouds whose internal pressure decouples from the volume-filling medium. 
        We verified that our results are not sensitive to the exact value of this cut.
        
        Figure~\ref{f:energy_summary} plots the ratios $\ephi//\etot/$ (top) and $\etot//E_{\rm grav}$ (bottom) for the three simulations, where $E_{\rm kin}=E_R+E_\phi+E_z$ is the total kinetic energy. 
        As above, we smooth the curves using a 100 Myr moving window for improved readability. 
        The figure shows that after \tbursty/, ISM energetics indicate a dynamically-cold rotating disc, with the rotational energy dominating the other kinetic ($R$ and $z$) and thermal energy components, and equalling roughly half the gravitational energy. 
        In contrast, prior to \tbursty/, rotational energy is subdominant, and the total kinetic + thermal terms exceed $0.5E_{\rm grav}$ with large dispersion on short time-scales, suggesting the galaxy is out-of equilibrium and unbound.
        This figure thus demonstrates that there is no rotation-dominated ISM before \tbursty/ in our simulations.
        
        Figures \ref{f:type_breakdown} and \ref{f:grav_ratio_type_breakdown} further explore ISM energetics as a function of $t-\tbursty/$, by plotting all kinetic and thermal components, and by partitioning the gas into different temperature phases: cold ($T<10^3$ K, second panels from the left), warm ($10^3 < T < 10^5$ K, third panels) and  hot ($T>10^5$ K, right-most panels). 
        We use \mtwelveb/ as an exemplar but the conclusions are valid also for \mtwelvei/ and \mtwelvef/. 
        The left panel of Figure~\ref{f:type_breakdown} demonstrates that prior to \tbursty/ the thermal energy and different kinetic energy components are comparable, with rapid short-timescale fluctuations.
        At $t\approx \tbursty/$ the ISM becomes rotation-dominated, as indicated by $E_\phi$ dominating other forms of energy.
        The three right panels show that the kinetic energy originates in the cold and warm phases, and that also in these phases there is no strongly preferred orientation prior to \tbursty/ while  rotation dominates after \tbursty/. 
        In the hot phase the total energy is dominated by thermal energy at all times, though its relative contribution to the total energy is small after \tbursty/ (see \S \ref{s:phase_breakdown} below). 
        One can also notice that in the hot phase $E_\phi$ increases relative to $E_R$ and $E_z$ at $\tbursty/$, suggesting a similar split in kinetic energy as in the cold and warm phases.
        
        The left panel of Figure~\ref{f:grav_ratio_type_breakdown} compares the different kinetic and thermal energy terms in \mtwelveb/ with $E_\mathrm{grav}$. 
        As suggested by Figure~\ref{f:energy_summary}, a dynamically cold disc with $\ephi/\approx E_{\rm grav}/2$ and $ E_R$, $E_z$, \etherm/ $\ll\ephi/$ is evident only after \tbursty/.
        Before \tbursty/ the energy in the four kinetic and thermal components are roughly equal and their sum exceeds $0.5 E_{\rm grav}$. 
        The three right panels show that the total kinetic energy in the cold and warm phases sums to $0.5 E_{\rm grav}$ also during the $\approx3$ Gyr prior to \tbursty/, suggesting a type of virial equilibrium in these phases. 
        At earlier times the kinetic energy exceeds $0.5E_{\rm grav}$ with large fluctuations, suggesting out of equilibrium dynamics. 
        In the hot phase (rightmost panel) the thermal energy exceeds $E_{\rm grav}$ at all times, indicating a thermally driven outflow (consistent with our previous findings at low-redshift, see Figure 11 of \paperone/), but by a larger factor before \tbursty/ suggesting stronger winds during the bursty phase.

    \subsection{The relative importance of different temperature phases in the ISM} \label{s:phase_breakdown}
        In this section we show the relative importance of the different temperature phases for the mass and energy of the ISM. 
        The top panel of Figure~\ref{f:phase_breakdown} plots the fraction of mass in each of the three different ISM temperature phases as a function of $t-\tbursty/$, for \mtwelveb/. 
        As above we smooth the curves with a 100 Myr window for readability.
        The figure shows that the warm phase dominates the mass budget prior to \tbursty/, with a subdominant but significant contribution from the cold phase and a typically small contribution from the hot phase. 
        The fraction of cold gas substantially increases past \tbursty/, becoming comparable to the warm phase, likely since higher densities and colder temperatures can be sustained after the ISM settles into a thin disc.
    
\figurenv{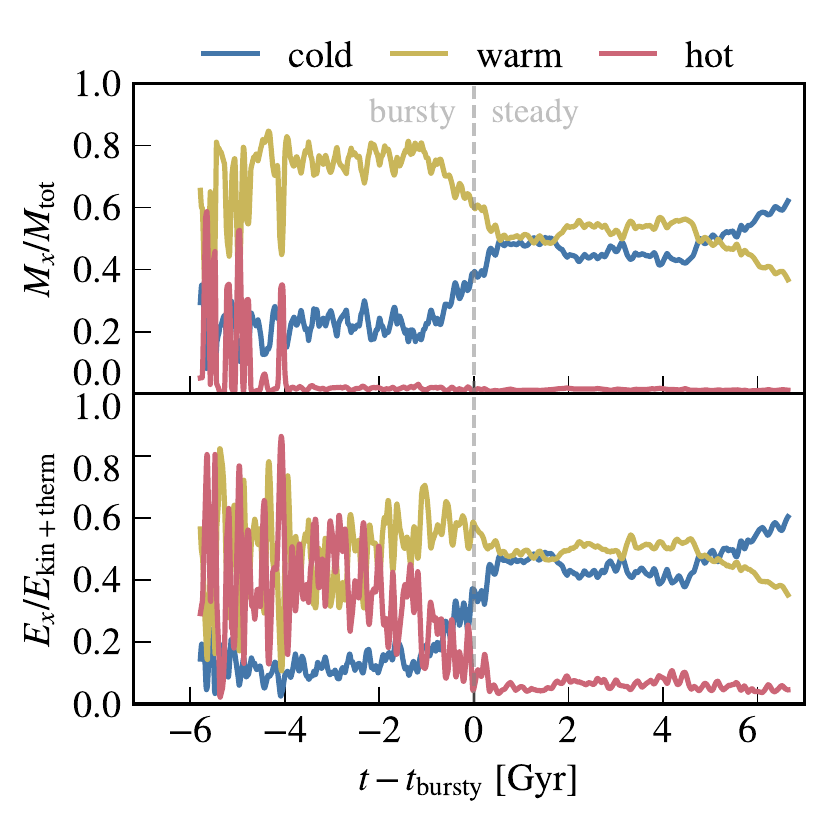}{
    \label{f:phase_breakdown} 
    The fraction of mass \textit{(top)} and energy \textit{(bottom)} in each of the cold ($T < 1000$ K), warm ($1000 < T < 10^{5}$ K) and hot ($T>10^5$ K) phases in the ISM of the \mtwelveb/ simulation. 
    At \tbursty/ the cold gas mass and energy fractions increase, while the hot gas energy fraction and variability decrease.}
    
        The bottom panel in Figure~\ref{f:phase_breakdown} plots the fraction of ISM energy in each temperature phase. 
        The hot and warm phases dominate the energy budget  before \tbursty/, with comparable contributions and large fluctuations. Around \tbursty/ both the contribution of the hot phase and the fluctuations substantially decrease, and the energy contribution from the cold phase becomes comparable to that of the warm phase. 
        Figure~\ref{f:phase_breakdown} thus supports the result from Figures~\ref{f:energy_summary}--\ref{f:grav_ratio_type_breakdown} that the ISM energetics go through a transition at $t\approx\tbursty/$. 

    \subsection{Applying these metrics to the inner CGM} \label{s:CGM}

\figurenv{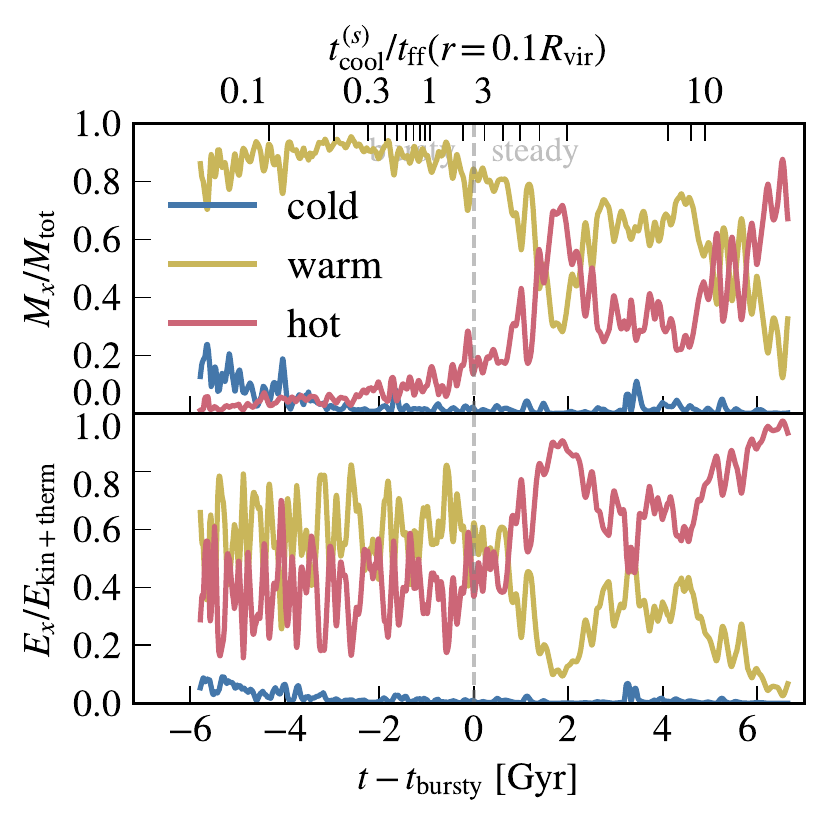}{
    \label{f:CGM_phase_breakdown}
    The fraction of mass \textit{(top)} and energy \textit{(bottom)} in each of the cold ($T < 1000$ K), warm ($1000 < T < 10^{5}$ K) and hot ($T>10^5$ K) phases, in a thin shell at $r=0.1\rvir/$ corresponding to the inner region of the CGM of the \mtwelveb/ simulation. 
    After \tbursty/, hot gas in the inner CGM dominates the energy and has a substantial contribution to the mass. Prior to \tbursty/ the hot CGM phase exhibits large short-timescale variability, is comparable in energy to the warm phase, and negligible in mass. 
    Comparison with Figure~\ref{f:phase_breakdown} demonstrates that the mass and energy distributions in the inner CGM are qualitatively similar to those in the ISM prior to \tbursty/, but differ strongly after \tbursty/. 
    The top axis denotes the ratio of the hot gas cooling time to the free-fall time. 
    The hot CGM is expected to become dominant and steady when this ratio exceeds $\sim1$, consistent with the trend seen in the plot.}
    
\figurenv{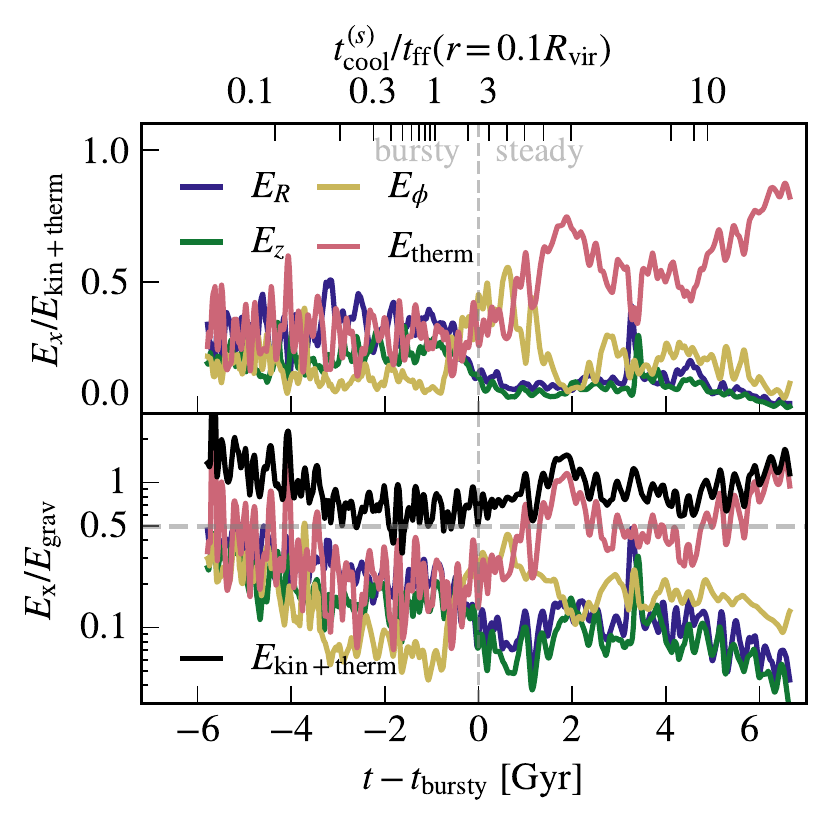}{
    \label{f:CGM_type_breakdown} 
    As Figure~\ref{f:type_breakdown}, but in a thin (0.1 dex) shell at $r=0.1\rvir/$ corresponding to the inner CGM. 
    Different colors denote the thermal component and the three kinetic components, while the black line in the bottom panel denotes the total thermal and kinetic energy. 
    The top axis denotes the ratio of the hot gas cooling time to the free-fall time. 
    After \tbursty/ thermal energy dominates. 
    Prior to \tbursty/ the inner CGM energy is split roughly equally between each of the kinetic components and the thermal energy, similar to the ISM prior to \tbursty/ (see Figure~\ref{f:type_breakdown}). 
    The ISM and inner CGM differ strongly in their distribution of energy components only after \tbursty/, with the ISM dominated by rotation and the inner CGM dominated by thermal energy. }

        \cite{Stern2020} showed that the inner CGM in FIRE `virializes' (i.e. becomes dominated by a steady,  thermal-pressure-supported hot phase) when the halo mass exceeds ${\sim}10^{12}$ M$_\odot$. 
        They also showed that this transition is associated with the transition from bursty to steady star-formation in the central galaxy and disc settling, potentially indicating a causal connection. 
        In this section, we first explore the process of inner CGM virialization with the same energy formalism employed above for the ISM, and then explore the connections between the transitions in the CGM and ISM. 
            
        Figure~\ref{f:CGM_phase_breakdown} plots the mass and energy fractions of the different temperature phases in the inner CGM, similar to the ISM analysis in Figure~\ref{f:phase_breakdown}.
        The bottom panel shows that prior to \tbursty/ the energy in the hot phase is highly variable and roughly equal to the energy in the warm phase, while the top panel shows that the mass is dominated by the warm phase. 
        In contrast, after \tbursty/ the hot phase dominates the energy and has a significant contribution to the mass, while the rapid fluctuations strongly decrease. 
        The contribution of the cold phase to the mass and energy is small in the CGM both before and after \tbursty/.
        The top panel of Figure~\ref{f:CGM_type_breakdown} partitions the energy in the inner CGM into its kinetic and thermal components.
        Prior to \tbursty/ most of the energy in the inner CGM is kinetic, while after \tbursty/ the thermal term is the largest, comprising more than half of the total energy. 
        The bottom panel shows that the thermal energy is roughly equal to the gravitational energy after \tbursty/ indicating that the thermal energy is sufficient to support the gas against gravity. 
        This transition, from an inner CGM dominated by kinetic motions to an inner CGM dominated by thermal pressure, is the virialization of the inner CGM identified by somewhat different methods in  \cite{Stern2020}.
    
        In idealized CGM calculations with and without stellar feedback, the CGM virializes when the cooling time of shocked hot gas $\tcools/$ exceeds the free-fall time $\tff$ \citep[e.g.,][]{WhiteRees78,birnboim2003, Fielding2017}. 
        \cite{Stern2020_hot_max} used analytic calculations and idealized simulations to argue that this is true in a local sense, i.e. that the inner CGM just outside the galaxy virializes when the local value of \tcooltff/ exceeds unity, a result later shown to hold also in the FIRE simulations \citep{Stern2020}. 
        Here we demonstrate this correspondence between \tcooltff/,  virialization, and the end of bursty star formation by noting $\tcooltff/(0.1\rvir/)$ in the top axes of Figures \ref{f:CGM_phase_breakdown} and \ref{f:CGM_type_breakdown}. 
        We calculate $\tff$ as \citep[e.g.,][]{McCourt12}
        \begin{equation}\label{e:tff}
            \tff=\frac{\sqrt{2}r}{v_c} ~,
        \end{equation}
        where $v_c=\sqrt{GM(<r)/r}$ and $M(<r)$ is the enclosed mass. 
        The value of $\tcools/$ is calculated as
        \begin{equation}\label{e:tcool}
            \tcools/\equiv t_{\rm cool}(T_c,P_{\rm HSE},Z,z)~,
        \end{equation}
        where $t_{\rm cool}$ is the cooling time based on the \cite{Wiersma2009} tables, $T_c\equiv0.6\mu m_{\rm p}v_c^2(r)/k_{\rm B}$ is comparable to the virial temperature, $Z$ is the shell-averaged metallicity in the snapshot, and $P_{\rm HSE}$ is the pressure expected in hydrostatic equilibrium, i.e.~it is equal to the weight of the overlying gas (equation 12 in \citealt{Stern2020}). 
        The value of $\tcools/$ thus approximates the cooling time in a hot, thermal pressure-supported CGM, regardless of whether such a virialized CGM actually exists in the snapshot, and is thus analogous to the cooling time estimate used in the idealized calculations mentioned above. 
        We refer the reader to \cite{Stern2020,Stern21b} for a discussion of $\tcools/$, how it evolves in the simulations as a function of mass and radius, and how it compares to other cooling time estimates. 
        We note also that calculating $\tcools/$ based on the average gas density in the shell rather than on $P_{\rm HSE}$ yields similar results. 

        Figures~\ref{f:CGM_phase_breakdown}--\ref{f:CGM_type_breakdown} demonstrate that the transitions in the CGM occur when \tcooltff/ exceeds $\approx2$, consistent with the order unity value expected from the idealized calculations mentioned above, and with the range of $\tcooltff/\sim1-4$ deduced in \cite{Stern2020}.  
        This result also demonstrates that in FIRE the epoch where \tcooltff/ exceeds unity and the inner CGM virializes are coincident with the epoch where star formation becomes steady and the ISM settles into a rotating disc.  
        In Appendix~\ref{a:tcool_tff} we show that this correspondence between $\tcools/\approx\tff$ and $t\approx\tbursty/$ applies to all three simulations analysed in this study. 
        This relationship has been found to hold also in additional FIRE galaxies by \citeauthor{Yu2021} (\citeyear{Yu2021}, see Figure~9 there).

\section{Discussion}\label{s:discussion}

    In this paper we use FIRE-2 cosmological zoom-in simulations to explore the properties of the ISM and inner CGM in MW-like galaxies versus time. 
    We demonstrate that these FIRE galaxies experience two qualitatively distinct phases, separated by a transition period which is short ($\lesssim 1$ Gyr) relative to cosmological timescales of $\sim5-10$ Gyr at the transition epoch ($z\sim 0.5-0.8$). 
    In the late phase, the star formation rate is steady, and the ISM forms a rotation-dominated disc with a narrow angular momentum distribution (Figure~\ref{f:diskiness}). 
    During this phase rotational energy comprises $>90\%$ of the total kinetic and thermal energy in the ISM and equals half the gravitational energy, as expected in a dynamically cold,  rotationally-supported disc (Figure~\ref{f:grav_ratio_type_breakdown}). 
    In contrast in the early phase, star formation is bursty (Figure~\ref{f:star_formation_history}), the geometry is quasi-spherical, and ISM energetics are dominated by bulk flows (including both turbulence and coherent inflows/outflows) with no clear preference for rotation (Figures~\ref{f:energy_summary}--\ref{f:type_breakdown}). Also, the ISM is on average hotter at these early times, with a smaller fraction of the ISM in the cold phase ($T<10^3$ K) than in the late disc phase, and with a hot phase which is more dynamically important (Fig.~\ref{f:phase_breakdown}). 
    The ISM thus transitions from being supported against gravity by quasi-isotropic flows prior to \tbursty/ to being supported by rotation after \tbursty/. 
    We further find that this transition in ISM properties is coincident with a transition in the inner CGM, from a medium supported by kinetic energy (dispersion and bulk flows) to a medium supported by thermal energy (Figures~\ref{f:CGM_phase_breakdown}--\ref{f:CGM_type_breakdown}). 

    In this section we discuss several implications of our results, and possible physical mechanisms which could be driving the transitions identified in the simulations. 

    \subsection{A challenge for standard models of high-redshift galaxies}
        In \paperone/, we analysed the same simulated galaxies as in this work, focusing on their late-time properties, well after disc settling. 
        In that study, we found that on average the total ISM pressure in the disc approximately balances the weight of the overlying gas (in the $z$ direction). 
        This finding provided support for equilibrium disc models, which have been used to explain the origin of the Kennicutt-Schmidt relation as arising from a balance between stellar feedback and gravity in the disc \citep[e.g.,][]{Thompson2005, Ostriker2011, Faucher-Giguere2013, Krumholz2018, Furlanetto2021_equilibrium}.
        In contrast, the results in this paper suggest that these equilibrium disc models do not apply to galaxies whose discs have not yet settled. 
        In the equilibrium models, it is assumed that the ISM is well modeled by a rotationally-supported disc, and that the vertical structure is determined by hydrostatic balance. 
        In the limit in which the ISM pressure is dominated by gas turbulence with a velocity dispersion $\sigma_{\rm g}$, the assumption of an equilibrium disc structure implies $\sigma_{\rm g} \lesssim v_c$ ($\sigma_{\rm g} \ll v_c$ for a thin disc), which in turn implies $E_{\rm kin+therm} \lesssim E_{\rm grav}$. 
        The results of \S \ref{s:energy_forms} show that the assumptions stated above are not satisfied before \tbursty/. 
        Rather, gas support is provided by quasi-isotropic bulk flows (including turbulence and coherent inflows and outflows powered by feedback and accretion).
        Moreover, values $E_{\rm kin+therm} > E_{\rm grav}$ imply that some of the gas is gravitationally unbound.

        It is sometimes assumed that the ``thick'' appearance of high-redshift galaxies can be explained by scaling relations derived for equilibrium discs supported by turbulence (e.g., $h/R \propto f_{\rm g}$). 
        Another common idea is that the massive bright ``clumps'' observed in high-redshift galaxies correspond to a Toomre mass ($M_{\rm T} \propto M_{\rm s} f_{\rm g}^{3}$), which is the characteristic mass for gravitational instability in discs \citep[e.g.,][]{Dekel2009b}. 
        These ideas have been useful to explain some properties of high-redshift galaxies found in observations and in simulations \citep[e.g.,][]{Genzel2011_clumps, Oklopcic2017, Dekel2021_clumps}. 
        While these scalings may be appropriate for massive high redshift galaxies whose discs have already settled (see \S~\ref{s:where_and_when} below), our analysis suggests that they do not accurately model the more common, lower-mass galaxies that should not have stable discs. 
        Going forward, our decomposition of the total gas into different thermodynamic phases and forms of energy in the early bursty phase could be used as a starting point to develop more accurate and detailed ISM models for these early lower mass galaxies. 
        Some recent studies have generalized equilibrium disc models to include oscillations around an equilibrium state, as may result from a time delay between star formation and stellar feedback \citep[e.g.,][]{Orr2020, Furlanetto2021_bursty}. 
        However, it is unclear to what degree this approach can accurately describe highly dynamic galaxies prior to disc settling, or whether a full non-equilibrium approach is needed.

    \subsection{When and where do we expect discs?}\label{s:where_and_when}
        In the analysis presented in this paper we focus on three simulations of Milky Way-mass galaxies, and show that their discs settle roughly concurrently with the transition to a thermal energy-supported inner CGM (`inner CGM virialization'). 
        The previous analysis of \cite{Stern2020} suggests that the correspondence between these transitions holds across a wide range of mass scales and redshifts, which allows us to extend the theoretical predictions to different mass scales.
        This is because the inner CGM is predicted to virialize at a halo mass $M_{\rm h}\sim 10^{12}$ M$_{\odot}$, roughly independent of redshift \citep[][]{Stern2020_hot_max, Stern2020, Stern21b}. 
        This predicts that halo mass, or equivalently the galaxy stellar mass given the weak evolution of the stellar mass-halo mass relation with redshift \citep[][]{Moster2018, Behroozi2019}, is a good proxy of whether a galaxy's disc is expected to have settled. 
        This implies that disc galaxies can exist at any redshift in sufficiently massive haloes, although discs will be relatively rare at high redshift due to the rarity of massive haloes. 
        This picture is qualitatively consistent with the recent systematic analysis of integral field data by \cite{Tiley2021}, which indicates that mass and not redshift is the primary variable determining the diskiness of $z\sim0-1.5$ galaxies. 
        Moreover, this would explain observations of rare large discs at higher redshifts, $z\gtrsim 4$ \citep[e.g.,][]{Neeleman2020,Rizzo2020,Tsukui2021}. 
        
        Our simulation results also appear broadly consistent with empirical results on the formation of the Milky Way. 
        Recently, Belokurov \& Kravtsov (in prep.) used abundance measurements from APOGEE and astrometry from Gaia and demonstrated that low-metallicity ($[{\rm Fe/H}]\lesssim -1.3$) stars formed in situ (i.e., not accreted from other galaxies) have an approximately isotropic velocity ellipsoid and relatively little net rotation. This stellar component, which they named `Aurora', suggests an early phase of quasi-isotropic evolution in the Milky Way prior to disc formation, similar to our results. 
        Belokurov \& Kravtsov further show that the tangential velocity of the in situ stars, i.e. the degree of rotational support, increases sharply with metallicity between $[{\rm Fe/H}]= -1.3$ and $[{\rm Fe/H}]=-0.9$. 
        Comparing their results with a set of simulations of Milky Way-mass galaxies, including publicly-available FIRE-2 simulations \citep[][]{Wetzel2022_FIRE2_public}, they deduce a rapid disc formation time scale of $\approx1-2$ Gyr, similar to what our simulations predict, though the Milky Way disc appears to form earlier and at lower metallicity than in the three simulations analysed in this paper. 
        Interestingly, Belokurov \& Kravtsov note that disc settling is accompanied by a qualitative change in chemical abundances (both in the observations and in the simulations), such that the scatter in abundances disappears around the time of disc settling.
        The larger scatter in stellar abundances prior to disc settling is interpreted as being likely caused by strongly time-variable inflow, outflow, and star formation rates in the bursty phase.

    \subsection{Is there a distinction between the ISM and inner CGM?}\label{s:ism_cgm_distinction}
\figurenv{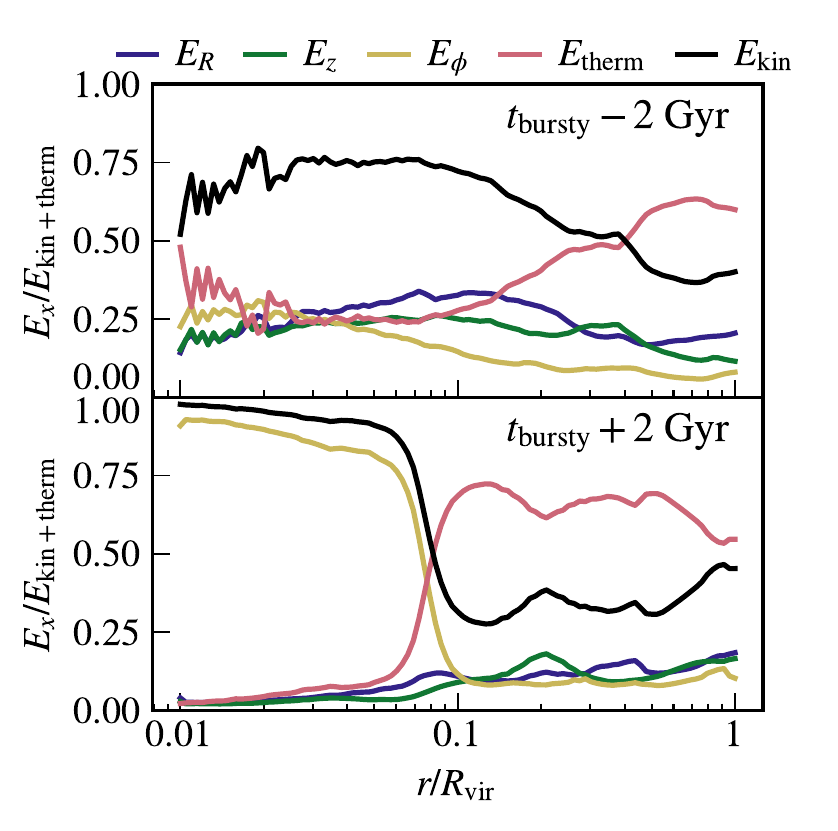}{
    \label{f:term_radial_profile} Ratios of different energy terms to the total kinetic+thermal energy as a function of radius in \mtwelveb/ at two different times: 2 Gyr before \tbursty/ (top) and 2 Gyr after \tbursty/ (bottom). 
    After \tbursty/ a clear transition is apparent between a rotation dominated ISM at ${\lesssim}0.08 \rvir/$ and a thermal energy dominated CGM at larger radii.
    In contrast, before \tbursty/, kinetic energy dominates with no preference for rotation at both ISM and inner CGM radii, out to ${\sim}0.3 \rvir/$.}

\begin{figure*}
\centering
\includegraphics[width=0.7\linewidth]{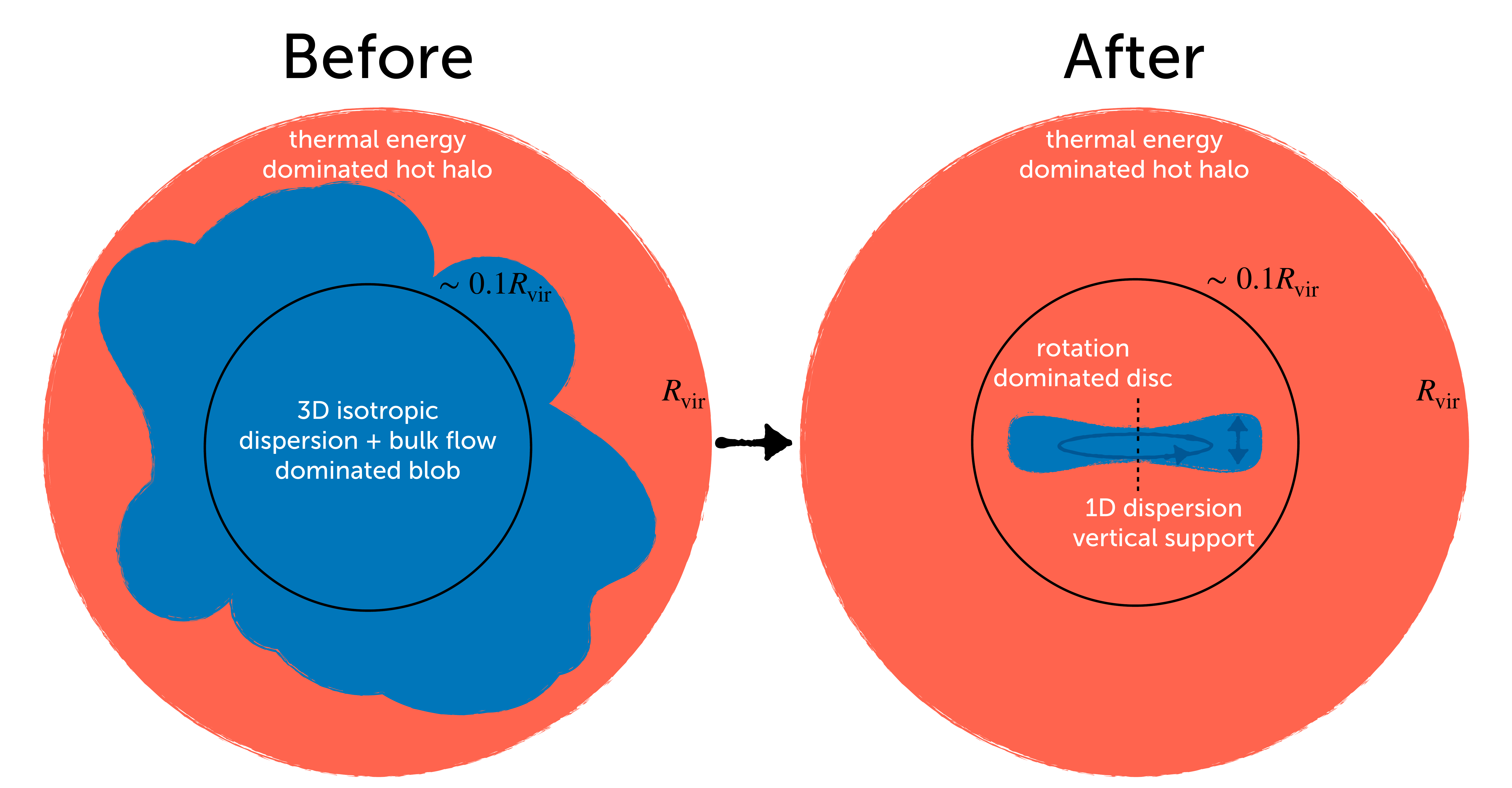}
\caption{\label{f:cartoon}
    A schematic diagram of gas energetics before and after \tbursty/, based on the quantitative results in Figure \ref{f:term_radial_profile} (not to scale).
    After \tbursty/ (right), there is a clear distinction between the rotation dominated ISM disc (with dispersion and bulk pressures supporting the disc vertically, see \paperone/) and the hot, thermal energy-supported CGM.
    This difference in rotational support naturally produces a clear boundary between hot CGM and cold ISM -- the hot CGM rapidly cools when rotational support is high.
    This is because the temperature of the hot, inflowing CGM is maintained by compressional heating as it moves inward. 
    This inward movement is drastically reduced when rotational support is large, leading to runaway cooling at the disk edge \citep{Hafen2022}. 
    In contrast, before \tbursty/ (left), there is a continuous medium dominated by turbulence and bulk in/outflows in all three directions, with no clear distinction between the energetics of the ISM and the inner CGM. 
    Red shows hot, virialized gas, and blue represents unvirialized gas (mostly cool, though warm/hot gas can be mixed in; see top row in Figure \ref{f:galaxy_stamps}).}
\end{figure*}

        The top row of Figure~\ref{f:galaxy_stamps} shows a complex morphology dominated by cool and warm gas extending to tens of kpc from the center at 2 Gyr before \tbursty/. 
        In contrast, in the second row of this Figure one can clearly identify cool thin discs embedded in hot surroundings at 2 Gyr after \tbursty/. 
        This visual distinction (or lack thereof) is supported quantitatively by comparing Figure~\ref{f:type_breakdown} with Figure~\ref{f:CGM_type_breakdown}, and is further explored in Figure~\ref{f:term_radial_profile} where we plot the different energy terms versus radius at $\pm2$ Gyr from \tbursty/. 
        The plotted energy terms are calculated in spherical shells with thickness 0.01 dex spanning from $0.01-1 \rvir/$  in each snapshot and averaged over a timespan $\pm$300 Myr to avoid short timescale fluctuations. 
        After \tbursty/ (bottom panel) the distribution of energy shows a sharp transition at $\approx0.08\rvir/$, from being entirely dominated by rotation at smaller radii (i.e.\ the ISM), to being entirely dominated by thermal energy at larger radii (the CGM). 
        In contrast, at the time before \tbursty/ shown in the top panel, kinetic energy dominates with $\eR/\sim\ephi/\sim\ez/$ at all radii out to $\approx0.3-0.4\rvir/$. 
        The top panel thus demonstrates that both in the ISM and in the inner CGM velocity dispersion and bulk inflows / outflows provide most of the support against gravity prior to \tbursty/, rather than rotation or thermal energy. 
        Figure~\ref{f:term_radial_profile} also demonstrates that this result is insensitive to the exact radius we use to attribute gas resolution elements to the ISM or inner CGM.
        
        Figures~\ref{f:phase_breakdown}--\ref{f:CGM_phase_breakdown} further demonstrate that before \tbursty/ the gas in both the ISM and inner CGM is dominated by the warm phase in terms of mass, while both the ISM and the inner CGM have comparable contributions from the warm and hot phases in terms of energy. 
        In contrast, after \tbursty/ the ISM is dominated by cold and warm phases while the inner CGM is hotter, dominated by the warm and hot phases. 
        It is thus only after disc settling and the virialization of the inner CGM that the `standard picture' for galaxy discs arises, in which a rotation-dominated, cool ISM is clearly distinct from a thermal pressure-dominated hot CGM. 
        Prior to these transitions both the ISM and the inner CGM are dominated by kinetic energy with no preferred orientation, and there is no clear distinction between their energetics. 
        The ISM and inner CGM form a single continuous medium supported by dispersion and bulk flows with intermediate temperatures.
        These two regimes are pictured schematically in Figure~\ref{f:cartoon}.
        
        The top panel of Figure~\ref{f:term_radial_profile} shows  that at large CGM radii ($>0.4\rvir/$) thermal energy dominates also before the transition at $t=\tbursty/-2$ Gyr.
        \cite{Stern2020} and \cite{Stern21b} showed this is likely a result of  $\tcooltff/$ increasing with radius and having values larger than unity in the outer CGM even when in the inner CGM this ratio is lower than unity and kinetic energy dominates. 
        Our results thus suggest that after \tbursty/ when $\tcools/ /\tff\gg1$ at all radii, the distinction between the CGM and ISM is a result of rotational support, which is large in the ISM and small in the CGM.
        In contrast, prior to \tbursty/ the distinction between the outer CGM and the inner CGM/ISM is a result of cooling, which is rapid at small radii and slow at large radii. 
        We note that the outer CGM also experiences a transition from being dominated by kinetic energy to being dominated by thermal energy, though at earlier times than this occurs in the inner CGM (Figure~12 in \citealt{Stern2020}). 
        At these very early times we expect the entire ISM and CGM system to be dominated by kinetic energy. 

    \subsection{What are the mechanisms that drive disc settling?}
        \newcommand{\vc}{v_{\rm c}}
        Above we find that at $t<\tbursty/$ the ISM is supported against gravity by turbulence and coherent flows rather than by rotation. 
        Given the rapid dissipation expected of such bulk flows ($\sim100$ Myr), the energy and momentum required to sustain them in our simulations is likely continuously supplied by a combination of accretion energy and stellar feedback.
        It thus seems that the mechanism which drives disc settling is a mechanism which changes the effects of stellar feedback, from a relatively strong mode of feedback which dominates rotation and hence entirely disrupts the disc, to a weaker feedback mode which is subdominant to rotation and hence provides support primarily in the vertical direction, as in standard equilibrium disc models. 
        Our results suggest that, for the Milky Way-mass galaxies we have analysed, this transition in the effects of stellar feedback occurs within a timescale of $\sim$ Gyr, as indicated by the transition time from $E_\phi/(E_{\rm kin}+E_{\rm therm})\approx0.2$ to $E_\phi/(E_{\rm kin}+E_{\rm th})\approx0.9$ in Figures~\ref{f:energy_summary}--\ref{f:grav_ratio_type_breakdown}. 
        This transition timescale is short compared to cosmological timescales of $\sim10$ Gyr, though it is long relative to galactic dynamical timescales of $\sim100$ Myr.
        
        Previous studies have shown that galaxy-scale outflows subside in FIRE when the SFR becomes steady \citep{Muratov2015, AA17_cycle, Pandya2021}.
        However, the physical mechanisms or properties driving these changes in SFR and feedback properties remain unclear. 
        Previous studies based on idealized models and arguments suggested that a qualitative change in stellar feedback-driven outflows is expected when the disc crosses a threshold in the depth of its gravitational potential well \citep{Dekel86}, in its gas fraction \citep{Hayward2017,Fielding2018,Orr2020}, in its dynamical time \citep{Torrey2017, Faucher-Giguere2018, Furlanetto2021_bursty}, or in its star formation rate surface density \citep{Murray2011_critical_Sigma, Kretschmer2020}. 
        All these quantities are correlated in our simulations, so it remains to be seen which of these suggested thresholds, if any, is the ultimate driver of the transitions. 
        We note that many of these idealized arguments are based on analytic derivations that assume there is an equilibrium disc to begin with, a condition which is not met in our simulations at early times, so some of the thresholds derived may not apply as predicted. 

        The rough concurrence between \tbursty/ and the transition to a thermal energy-supported inner CGM (Figures~\ref{f:CGM_phase_breakdown}--\ref{f:CGM_type_breakdown}) and the similarity in the energetics of the ISM and inner CGM prior to \tbursty/ (Figures~\ref{f:term_radial_profile} -- \ref{f:cartoon}) suggest that the process of `inner CGM virialization' plays an important role in disc settling, as proposed by \cite{Stern2020}. 
        Inner CGM virialization causes accreted gas to have a narrow angular momentum distribution prior to accretion onto the central galaxy, due to more efficient angular momentum exchange in a thermal energy-dominated CGM, in contrast with a large dispersion in angular momentum in a cool CGM dominated by kinetic energy \citep{Hafen2022}. 
        A thermal energy-dominated CGM is also more efficient in confining galactic outflows due to its uniformly high pressure, in contrast with large pressure fluctuations when kinetic energy dominates, which allow outflows to expand through paths of least resistance \citep[][]{Stern2020}.\footnote{Some authors have previously suggested that stellar feedback-driven outflows may be suppressed after the CGM virializes because the outflows lose buoyancy when the halo becomes filled with hot gas \citep[][]{Bower17, Keller2020_buoyancy}. 
        The effect mentioned here is different in that it focuses on the effects of pressure fluctuations in creating paths of least resistance in the CGM.} 
        These transitions in accretion and feedback properties following inner CGM virialization would both be conducive to the formation of steady discs and a decrease in ISM velocity dispersion. 
        
        An important aspect of inner CGM virialization is that it can potentially explain in simple halo-based terms when, during their cosmological evolution, galaxies develop stable discs. 
        This is because CGM virialization is expected to occur when \tcooltff/ exceeds unity, even when galactic outflows are neglected and insensitive to the detailed properties of the central galaxy. 
        Indeed, a large number of cosmological simulations that neglected outflows and treated the ISM in subgrid found that gaseous haloes become hot when they reach a total mass $\sim {\rm few}\times 10^{11}$ M$_{\odot}$ \citep[e.g.,][]{Keres05, Keres2009a, Faucher-Giguere2011a}. 
        As mentioned in \S \ref{s:where_and_when}, in reality the critical halo mass for virialization depends e.g. on gas metallicity, which is sensitive to feedback. 
        However, for a range of well-motivated assumptions, analytic scalings and idealized models predict that the CGM should complete virialization within roughly $\pm 0.5$ dex in halo mass from what is predicted neglecting feedback and metals \citep{birnboim2003, Stern2020_hot_max}, and this is indeed what is found in a larger sample of FIRE simulations that include realistic stellar feedback and metal enrichment \citep[][]{Stern2020}. 
        
        A possible alternative interpretation to the coincident inner CGM virialization and disc settling would be the reverse causality, i.e. that disc settling causes the inner CGM to virialize. 
        For example, this could happen if disc settling causes the outflows to be suppressed due to physical changes internal to the galaxy \citep[e.g. if changes in the density PDF of the ISM make outflows less efficient at escaping through low-density channels;][]{Hayward2017}.  
        The suppression of galactic outflows could cause gas densities and metallicities in the inner CGM to decrease, which would increase $t_{\rm cool}$ and could push the gas toward a virialized stated. 
        While this cannot be ruled out based on our analysis, this would require disc settling to somehow occur around the same halo mass scale where we expect CGM virialization to occur without interacting with feedback.
        Alternatively, inner CGM virialization and disc settling could be concurrent by coincidence due to correlations between \tcooltff/ and other properties such as $\vc$.

        Finally, we note that even if CGM virialization is a primary driver of disc settling, there could be mutually reinforcing effects involved. 
        For instance, if inner CGM virialization initiates disc settling and contributes to suppressing outflows, then \tcooltff/ could increase for the reasons above, thus accelerating the virialization process. 
        As a disc settles, its decreasing gas fraction and thickness (see Figures \ref{f:basic_properties} and \ref{f:diskiness}) bring down the Toomre mass $M_{\rm T}\propto M_{\rm s} f_{\rm g}^{3}\propto \Sigma_{\rm g} h^{2}$. 
        A smaller Toomre mass implies a larger number of lower-mass star-forming regions distributed throughout the galaxy (as opposed to a small number of dominant clumps) and may reduce the disruptive effects of stellar feedback by limiting the clustering of supernovae. 
        Mutually-reinforcing effects could help explain the rapidity of disc settling once the process starts.
        
        Future work should attempt to more stringently test the causal links between inner CGM virialization and disc settling. 

\section{Conclusions} \label{s:conclusion}
    We analyse three FIRE-2 cosmological zoom-in simulations of Milky-Way mass galaxies in order to characterize the properties of their ISM and inner CGM before and after `disc settling.'  
    This analysis has implications for our theoretical understanding of the formation of disk galaxies, and may also be important to understand recent the disc settling phenomenon highlighted by recent morphological and kinematic observations of star-forming galaxies at range of redshifts \citep[e.g.,][]{Kassin2012a, vanderWel14, Simons2017,Tiley2021}. 
    We find the following:
    \begin{enumerate}
        \item Milky Way-mass galaxies in the FIRE-2 simulations experience two phases of galaxy growth, separated at $z\sim 0.5 -0.8$: a dynamic ``bursty'' phase of star formation at early times and a stable ``time-steady'' phase of star formation at late times (Figures~\ref{f:galaxy_stamps}--\ref{f:star_formation_history}). 
        This transition between order-of-magnitude different levels of star formation variability is consistent with previous analyses of FIRE simulations \citep{Muratov2015, AA17_cycle, Sparre2017, Faucher-Giguere2018}.
        \item The transition from bursty to time-steady star formation is concurrent with the emergence of a rotationally-supported gaseous disc (Figures~\ref{f:diskiness}--\ref{f:grav_ratio_type_breakdown}). 
        Prior to the transition the ISM is instead supported against gravity by quasi-isotropic bulk flows, including turbulence and coherent inflows/outflows. 
        Moreover, at early times the total thermal + kinetic energy in the gas frequently exceeds the gravitational binding energy by large factors. 
        These results challenge the applicability of standard equilibrium disc models \citep[e.g.,][]{Faucher-Giguere2013, Krumholz2018} to high-redshift galaxies prior to disc settling. 
        A quasi-isotropic ISM in the progenitors of Milky Way-like galaxies is consistent with the isotropic velocity distribution of low-metallicity stars formed in situ in the Milky Way (Belokurov \& Kravtsov, in prep.).
        \item The emergence of a rotation-dominated ISM in FIRE occurs over a ${\sim}$Gyr timescale, which is short relative to cosmological timescales of $5-10$ Gyr at the transition epoch (e.g., Figure~\ref{f:type_breakdown}). 
        % The rapidity of disc settling predicted by the simulations is consistent with the rapid emergence of a rotationally-supported disc in the Milky Way indicated by a sharp increase of the tangential velocity component with increasing metallicity for stars formed in situ (Belokurov \& Kravtsov, in prep.; private communication).
        \item We find that the transition to a rotation-supported ISM with steady star formation is concurrent with a transition to thermal energy support at small CGM radii (`inner CGM virialization', Figures~\ref{f:CGM_phase_breakdown}--\ref{f:CGM_type_breakdown}), consistent with the results of \cite{Stern2020}. 
        This implies that prior to disc settling both the ISM and inner CGM are supported by bulk flows and hence it is not straightforward to distinguish between the ISM and the inner CGM, in contrast with after disc settling where angular momentum support provides a boundary. 
        This result also raises the possibility that the process of CGM virialization, discussed in the literature for over four decades \citep[e.g.,][]{rees77,Silk77,  WhiteRees78,WhiteFrenk91,birnboim2003,Keres05,Nelson13, Fielding2017,Stern2020_hot_max} may play a key role in driving the settling of galactic discs.
    \end{enumerate}
    There are several interesting directions for future work. 
    In this paper, we have presented an in-depth analysis of disc settling in simulated Milky Way-mass galaxies. 
    Going forward, it will be important to expand the analysis by applying the diagnostics developed herein to a larger sample of galaxies covering a wide range of halo mass. 
    This would allow us to quantify to what extent disc settling proceeds in the same way for galaxies with different growth histories.
    It will also be valuable to produce observational predictions that can be directly compared with observations in order to assess whether the phenomenon of disc settling apparent in observations matches the evolution that we find in FIRE around the transition from bursty to steady star formation. 
    Finally, there is more work to be done to disentangle causal links and more firmly establish the role of different physical factors, including inner CGM virialization, in driving the evolution of galaxy properties during disc settling. 
    Controlled numerical simulations, in which physical parameters such as galaxy properties vs. CGM properties can be independently varied (as opposed to cosmological simulations in which various correlations are built in), could be very useful to clarify causality.

\section*{Acknowledgements}
We thank Vasily Belokurov and Andrey Kravtsov for sharing an advance copy of their paper on the identification of the Aurora stellar component and its interpretation in terms of an early chaotic phase in the Milky Way's evolution. 
ABG was supported by an NSF-GRFP under grant DGE-1842165  and was additionally supported by NSF grants DGE-0948017 and DGE-145000.
JS was supported by the Israel Science  Foundation  (grant  No.~2584/21) and by the German Science Foundation via DIP  grant STE 1869/2-1 GE625/17-1. 
CAFG was supported by NSF through grants AST-1715216, AST-2108230,  and CAREER award AST-1652522; by NASA through grant 17-ATP17-0067; by STScI through grant HST-AR-16124.001-A; and by the Research Corporation for Science Advancement through a Cottrell Scholar Award.
Support for PFH was provided by NSF Research Grants 1911233 \&\ 20009234, NSF CAREER grant 1455342, NASA grants 80NSSC18K0562, HST-AR-15800.001-A. 
AW received support from: NSF grants CAREER 2045928 and 2107772; NASA ATP grants 80NSSC18K1097 and 80NSSC20K0513; HST grants AR-15809 and GO-15902 from STScI; a Scialog Award from the Heising-Simons Foundation; and a Hellman Fellowship.
AJR was supported by a COFUND/Durham Junior Research Fellowship under EU grant 609412; and by the Science and Technology Facilities Council [ST/T000244/1].
Numerical calculations were run on the Caltech computer cluster Wheeler, the Northwestern computer cluster Quest,
Frontera allocation FTA-Hopkins/AST20016 supported by the NSF and TACC, XSEDE allocations ACI-1548562, TG-AST140023, and TG-AST140064, and NASA HEC allocations SMD-16-7561, SMD-17-1204, and SMD-16-7592.
ZH was supported by a Gary A. McCue postdoctoral fellowship at UC Irvine.
The data used in this work were, in part, hosted on facilities supported by the Scientific Computing Core at the Flatiron Institute, a division of the Simons Foundation.

\section*{Data Availability}
The data supporting the plots within this article are available on reasonable request to the corresponding author. 
A public version of the GIZMO code is available at \url{http://www.tapir.caltech.edu/~phopkins/Site/GIZMO.html}.
FIRE-2 simulations are publicly available \citep{Wetzel2022_FIRE2_public} at \url{http://flathub.flatironinstitute.org/fire}.
Additional data including simulation snapshots, initial conditions, and derived data products are available at \url{http://fire.northwestern.edu/data/}.

%%%%%%%%%%%%%%%%%%%%%%%%%%%%%%%%%%%%%%%%%%%%%%%%%%



\bibliographystyle{mnras}
\bibliography{bibliography} % if your bibtex file is called example.bib

%%%%%%%%%%%%%%%%%%%%%%%%%%%%%%%%%%%%%%%%%%%%%%%%%%

%%%%%%%%%%%%%%%%% APPENDICES %%%%%%%%%%%%%%%%%%%%%

\appendix

\section{The effects of magnetic fields and cosmic rays}\label{a:cosmic_ray}
\figurenv{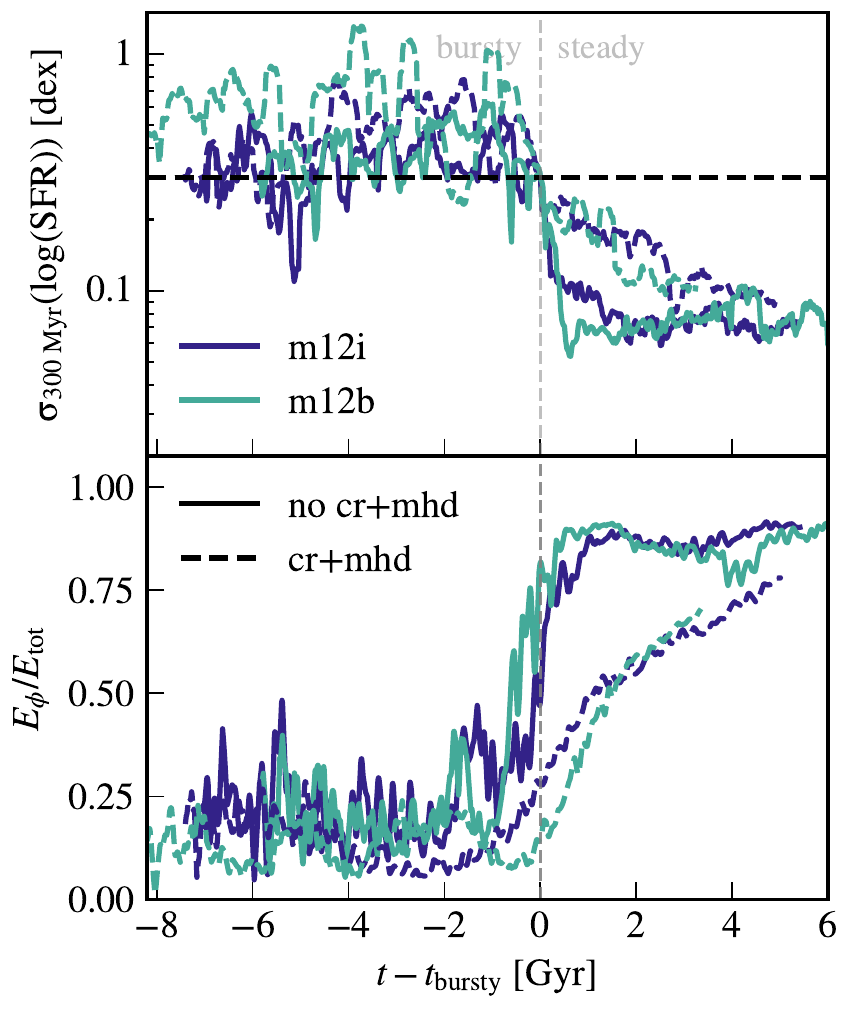}{
    \label{f:CR_SFR_diskiness} 
    Comparison of some basic results for two simulations with identical initial conditions, which include magnetic fields and either exclude  (solid) or include (dashed) cosmic ray physics.
    Top panel shows the running scatter in the SFR in a 300 Myr moving window, similar to Figure~\ref{f:star_formation_history}. 
    The bottom panel shows the ratio of rotational energy to total kinetic and thermal energy, as in Figure~\ref{f:energy_summary}.} 
    
    Figure~\ref{f:CR_SFR_diskiness} demonstrates the dependence of some of our results on the inclusion magnetohydrodynamics (MHD) and cosmic rays in our simulations, focusing the \mtwelvei/ and \mtwelvei/ runs, for which we have variants with and without these physics included. 
    The runs with MHD also include anisotropic thermal conduction and viscosity, while the runs with CRs include all these physics plus anisotropic diffusion and streaming of cosmic rays with an effective diffusion coefficient $\kappa=3\times10^{29}$ cm$^{2}$ s$^{-1}$. 
    These runs are labeled `MHD+' and `CR+' in previous papers that introduced them and analysed them in more detail \citep[e.g.,][]{Hopkins2020_what_about, Chan2021}. 
    
    The top panel of Figure~\ref{f:CR_SFR_diskiness} shows that there is a bursty SFR transition even when these physics are included. 
    The bottom panel shows that the fraction of kinetic energy, in the $\phi$-direction sharply increases at \tbursty/ in the MHD-only simulations as in the fiducial simulations analysed in the main text. 
    A more gradual increase in the fraction of rotational energy is seen in the simulations which include cosmic rays. 
    We defer exploring the effects of CRs in more detail to future work; these are most likely dependent on the (uncertain) assumptions adopted for cosmic-ray transport.

\section{Cooling time to free-fall time for each simulation}\label{a:tcool_tff}
\figurenv{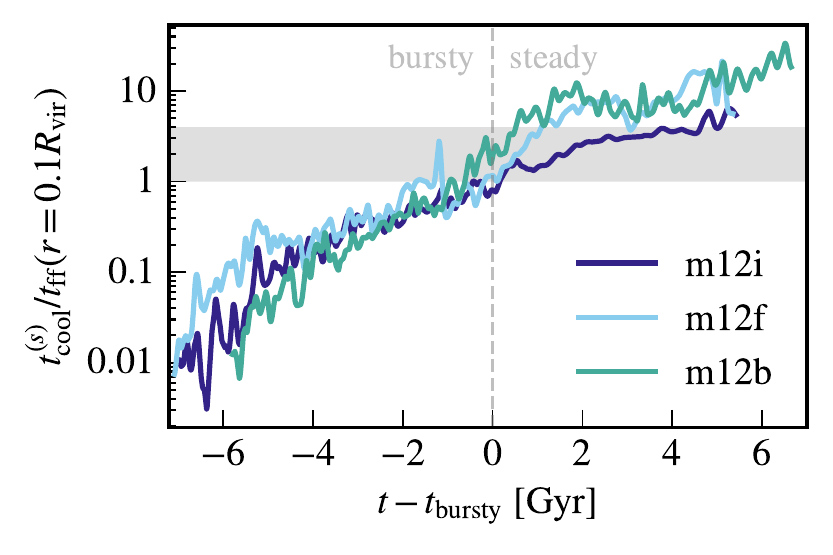}{
    \label{f:tcool_tff} 
    The ratio of the cooling time averaged over a thin shell at $0.1\rvir/$ to the free-fall time versus cosmic time offset by \tbursty/.
    The virialization of the inner CGM which occurs when $\tcooltff/ \approx1-4$ \citep{Stern2020} roughly coincides with \tbursty/ and the ISM transitions identified in this paper.
    }

    Figure~\ref{f:tcool_tff} shows the ratio $\tcools//\tff$ at $0.1\rvir/$, defined in \citet{Stern2020} and in \S \ref{s:CGM}, as a function of cosmic time.
    At late times when $\tcools/\gg\tff$ the inner CGM is dominated by a hot phase and is predominantly supported by thermal pressure, while at early times when $\tcools/\ll\tff$ the warm phase and kinetic pressure dominate (Figures~\ref{f:CGM_phase_breakdown}--\ref{f:CGM_type_breakdown}). 
    Comparing the intersection of the curves with the vertical line plotted at \tbursty/ reveals that inner CGM virialization coincides with the transitions we identify in the ISM, for all three simulations in our sample. 
    The horizontal gray band in Figure ~\ref{f:tcool_tff} corresponds to the range $\tcooltff/ = 1-4$. 
    The simple analytic expectation is that the CGM virializes when $\tcooltff/$ is of order unity, but \citet{Stern2020} finds that for our exact definition of this ratio, the inner CGM in the FIRE simulations typically virializes when $\tcooltff/$ is in this range (see \S 3.2 in \citeauthor{Stern2020}).

%If you want to present additional material which would interrupt the flow of the main paper, it can be placed in an Appendix which appears after the list of references.

%%%%%%%%%%%%%%%%%%%%%%%%%%%%%%%%%%%%%%%%%%%%%%%%%%

% Don't change these lines
\bsp	% typesetting comment
\label{lastpage}
\end{document}